%% file: templateArxiv.tex
\documentclass{article}
\usepackage{PRIMEarxiv}

\usepackage[utf8]{inputenc} 
\usepackage[T1]{fontenc}    
\usepackage{hyperref}       
\usepackage{url}            
\usepackage{booktabs}       
\usepackage{amsfonts}       
\usepackage{nicefrac}       
\usepackage{microtype}      
\usepackage{lipsum}
\usepackage{bm} 
\usepackage{fancyhdr}       
\usepackage{graphicx}       
\graphicspath{{media/}}     
\usepackage{natbib}
\usepackage{amsmath}
\usepackage{amssymb}
\usepackage{algorithm}
\usepackage{algpseudocode}
\usepackage{tikz}
\usepackage{threeparttable}
\usepackage{siunitx}
\usepackage{hyperref}
\usetikzlibrary{arrows.meta, positioning, calc, fit, backgrounds, decorations.pathreplacing}
\usetikzlibrary{external}

\usepackage{amsthm}
\newtheorem{definition}{Definition}
\newtheorem{proposition}{Proposition}

\title{Reproducing Fine-Grained Intervention Effects in High-Level Urban Air Quality Models: Abstraction Admissibility, Construction and Identifiability
}

\author{
  Xiang Xie, Manuel Herrera, Sheen Cabaneros, Asha Palihakkara, Philip James, Mohamad Kassem \\
  Newcastle University \\
  Newcastle upon Tyne\\
  \parbox{\textwidth}{\centering \texttt{\{xiang.xie, manuel.herrera, sheen.cabaneros, asha.palihakkara, philip.james, mohamad.kassem\}@newcastle.ac.uk}} \\
}

\begin{document}
\maketitle

\begin{abstract}
Urban air quality management operates on coarse aggregates, such as zones and vehicle fleets, rather than resolved atmospheric physics. While data-driven models reproduce monitored concentrations, they lack the causal mechanics required to infer unobserved emissions or predict the impact of novel interventions. This paper formulates a high-level urban air quality model as a causal abstraction of fine-grained physics, and derives the conditions under which coarse variables inherit the causal relations of the fine-grained models. First, abstraction over space and time is bounded by the transport physics: reporting intervals must be long enough for collapsed processes to reach local equilibrium, and short enough that spatial influence does not breach unit boundaries under steady transport. Second, the causal diagram recording the relations among the coarse variables is determined by the chemical mechanism and the spatial geometry, and each relation is estimated by reconstructing the observations. The fitted model answers observational queries by recovering unobserved quantities the model represents, and interventional queries by replacing one relation with an imposed value and evaluating the response. Third, an identifiability check determines whether an interventional query has a reliable estimate under the observations. In a controlled synthetic experiment, unrecorded construction activity confounds urban NO\textsubscript{x} and PM\textsubscript{2.5}, raising both simultaneously. The confounder is detected via the residuals of the confounder-unaware model. The fitted confounder-aware model estimates the fall in urban NO\textsubscript{x} and PM\textsubscript{2.5} that a 20\% traffic-emissions cut produces, and declines a non-identifiable query. The approach establishes a warranted paradigm for high-level air quality modelling: its answers remain grounded in represented physics, specific to the city, sources, and observations in question, and bounded by what the model can and cannot determine.
\end{abstract}

\keywords{Urban air quality \and Causal abstraction \and Neural causal model \and Abduction \and Intervention \and Neuro-symbolic AI
}

\section{Introduction}
\label{sec:Intro}
Human reasoning about the urban atmosphere is usually conducted over coarse variables. A planner does not reason about the instantaneous velocity and concentration fields resolved at the scale of a vehicle wake. Instead, the planner reasons about a low emission zone boundary, a fleet composition, a speed limit, and an annual-mean concentration at a receptor. This is not a limitation to be engineered away but a structural feature of the problem: the state of a resolved urban atmosphere is not representable at city scale, and every quantity that appears in an assessment, e.g., a grid-cell concentration, a fleet-average emission factor, a sector total, is already an aggregate of processes the model does not resolve. The question is whether a given aggregate supports the inferences drawn from it. A coarse variable (a.k.a. macro-variables) earns its place when relations among coarse variables track relations among the underlying fine ones (a.k.a. micro-variables) under the operations the model represents, which for air quality means emission controls, traffic measures, and fleet change. Recent work on causal abstraction defines this precisely: a high-level model is a valid abstraction of a low-level one when the map between them commutes with intervention, and the constructive counterpart derives macro-variables from micro-variables by requiring exactly that they preserve the causal relation \citep{chalupka2015visual, rubenstein2017causal, beckers2019abstracting}.

A common misunderstanding is that predictive fit establishes causal content. A high-level model that reproduces monitored concentrations well is licensed to describe the conditions under which it was fitted, but not to infer the quantities it does not observe, and still less to estimate the effect of a control intervention. Because such a model learns associations that hold under the prevailing regime, not the mechanisms that generate them. This problem has largely reshaped contemporary artificial intelligence, motivating work on the identifiability of causal representations \citep{scholkopf2021toward}, on testing whether an assumed high-level description is a causal abstraction of the network implementing it \citep{geiger2021causal}, and on neural models constrained to respect a causal structure \citep{xia2021causal}. A feasible solution that has proved most tractable is architectural rather than statistical, and belongs to the neuro-symbolic tradition of combining learned function approximation with explicit symbolic structure \citep{garcez2023neurosymbolic}: a symbolic object constrains what the neural component may represent and determines, prior to fitting, which queries are answerable at all. The neural causal model (NCM) is an instance of this arrangement, in which neural networks parameterise the mechanisms while the causal diagram constrains their dependence structure and supplies the graphical criterion by which a query is determined to be identifiable \citep{xia2021causal}.

High-level urban air quality models mainly serve three purposes in practice:
\begin{itemize}
    \item \textbf{Tractability}. Full chemical transport simulations are computationally expensive. Three-dimensional models such as CMAQ, WRF-Chem, and EMEP therefore replace near-explicit gas-phase chemistry with reduced mechanisms, and high-level models are layered above them when the application demands more or faster runs, e.g., operational forecasting, interactive scenario screening, and optimisation of control measures. 
    \item \textbf{Abduction}. A city is instrumented by a small number of stations supplemented by a partially reliable sensor network. However, practitioners may need estimates of emission strengths on unmonitored links, and of quantities that are not directly measurable, such as the prevailing NOx/VOC regime that determines whether ozone responds to a reduction in one precursor or the other.
    \item \textbf{Intervention}. Interventions like low-emission and clean-air zones, retrofit programmes and fleet turnover are defined at the coarse level. A model used to assess them must therefore get the causal relations among those aggregate variables right. This is not guaranteed: multiple high-level models may fit the available observations equally well, yet disagree on the effect of the same intervention. Whether a given interventional query is identifiable, or merely appears to be, depends on the abstraction and the observations, and should be established before answering.
\end{itemize}

\section{Literature Review}
\label{sec:review}
Operational urban air quality modelling is inherently a stack of abstractions. At the finest tier, computational fluid dynamics resolves the flow and concentration fields within individual street canyons, capturing the recirculating vortex and the roof-level exchange that controls kerbside concentrations \citep{sin2023cfd}. Above it, chemical transport models such as CMAQ, WRF-Chem, EMEP, and CHIMERE discretise the advection-diffusion-reaction equations on a grid of kilometre-to-tens-of-kilometre cells and integrate a reduced gas-phase mechanism \citep{xie2024improvement}. Dispersion models such as ADMS-Urban, AERMOD, and OSPM represent sources by Gaussian plumes or canyon parameterisations. Statistical and machine-learning tiers, such as land use regression, graph neural networks (GNNs) over station adjacency, and Gaussian-process or neural emulators of the physics, sit above these and are trained on outputs or observations \citep{han2022semi}. Each tier abstracts the one below it. Abstraction here is for tractability rather than a modelling convenience: the degrees of freedom of a resolved urban atmosphere cannot be carried at city scale, thus some coarsening is forced before any model can be written.

The review is organised by abstraction operation. Models are assessed against the abstraction operations they embody, focusing on whether what results remains capable of the purposes it is put to, in particular abduction and intervention. 

\subsection{Abstraction over the state space}
Spatial coarsening ($\tau_{space}$) and temporal aggregation ($\tau_{time}$) are the most common abstraction operations. $\tau_{space}$ partitions the domain into cells, each carrying a single value that represents the resolved field throughout its extent. A cell may be a grid square of a chemical transport model, a neighbourhood, or the catchment implicitly assigned to a monitoring station. $\tau_{time}$ replaces the sequence of states within a reporting interval with a single value, aggregating hourly outputs to daily, seasonal, or annual means. Land use regression, Gaussian-plume and dispersion-kernel models, graph neural networks over cell adjacency, and the aggregation stages of operational reporting all rest on them \citep{beelen2013development,wang2020pm2}. Spatial coarsening marginalises within-cell heterogeneity: a kerbside and a courtyard in the same cell respond differently to an identical fleet change, but a single cell value cannot represent both interventional responses. Temporal aggregation marginalises over meteorological states with different active edge sets, causing the annual mean to superimpose stagnant episodes, when local sources dominate, with ventilated periods, when they do not.

Neither spatial coarsening nor temporal aggregation is sound or unsound in itself. Whether it preserves the relations it is meant to represent depends on how the chosen scales relate to the transport and to the processes occurring within them. For example, a GNN's adjacency is fixed a priori, typically by geographic distance or correlation, whereas the set of cells that can influence a target within one time step is set by the advective range, which varies with the wind. The declared neighbourhood is thus a spatial abstraction chosen independently of the transport it stands for, and it is misspecified whenever the wind carries influence beyond the assumed neighbours \citep{wang2020pm2}. Such a mismatch is not apparent from model performance, since the misspecification is absorbed into the fitted parameters.

\subsection{Abstraction over composition}
Chemical lumping ($\tau_{species}$) and source aggregation ($\tau_{source}$) coarsen the composition of the mixture. Explicit VOC speciation is reduced to reactivity classes and ultimately to a single VOC variable, NO and NO\textsubscript{2} are combined into NO\textsubscript{x}, particulate components are combined into total mass, while individual vehicles are aggregated to a fleet and then to ``traffic''. These operations occur through mechanism reduction in chemical transport models, source-apportionment methods such as positive matrix factorisation and chemical mass balance, and emission-inventory construction. 

Chemical lumping is valid only where the response is invariant to the within-lump composition. This fails for VOC reactivity, because two mixtures with the same total VOC but different reactivity produce different ozone. In a high-NO\textsubscript{x} urban core, computed daytime ozone production rates diverge substantially across mechanisms, for example, 31.0, 29.6, and 25.0~$\text{ppbv}\cdot\text{h}^{-1}$ for RACM2, SAPRC07, and CB06, respectively, in a recent intercomparison, with a further study attributing a 20\% higher production rate in the explicit MCM~v3.3.1 relative to CB6r2 largely to the speciation of VOCs and organic peroxy radicals \citep{luecken2019multipollutant}. It fails for particulate matter because nitrate formation is a non-linear function of ammonia availability and aerosol pH, and equal-mass aerosols of different composition respond differently to the same precursor cut \citep{nenes2020aerosol, guo2018effectiveness}.

Source aggregation fails in different ways. A transport source combines diesel goods vehicles, petrol cars, buses, motorcycles, and rail, which differ in emission composition and magnitude, and thus the value assigned to it is a weighted average whose weights reflect the fleet composition at the time the inventory was compiled. Because many fleet compositions correspond to the same aggregate, an instruction expressed in terms of the aggregate does not determine what is to be changed. Reducing transport emissions by 20\% can be achieved by restricting goods vehicles during the morning peak or by reducing emissions across the fleet, and these produce different concentration fields. The instruction is therefore incomplete as an intervention specification, a difficulty identified in general terms as ambiguous manipulation \citep{spirtes2004causal}.

\subsection{Abstraction over driving conditions}
Meteorological reduction ($\tau_{met}$) replaces the resolved meteorological state with a small set of derived descriptors. The full state, e.g., the three-dimensional wind, temperature, and humidity fields, the turbulence structure, and the radiation field, is neither available at the coarse-model resolution nor compact enough to serve as a driver. Therefore, high-level models are driven instead by quantities such as a ventilation coefficient formed from wind speed and mixing height, a stability class, a dilution or dispersion rate, or a single effective wind. Dispersion models do this by parameterising turbulence with a stability category, statistical models use a handful of meteorological covariates as regressors, and surrogates use a reduced index rather than a full meteorological field.

The reduction is efficient because the descriptors are constructed to predict concentration. It is problematic because a single meteorological variable does not reach a pollutant along a single path. For instance, higher temperatures increase biogenic and evaporative VOC emissions and tend to increase ozone. It also accelerates reactions that form secondary pollutants, which act in the same direction. Meanwhile, it promotes the thermal decomposition of PAN, releasing NO\textsubscript{x} far from where it was emitted, and it drives convective mixing that deepens the boundary layer and dilutes what is present, which acts in the opposite direction. Wind speed behaves similarly: it dilutes local emissions, but it also increases road-dust resuspension and determines which upwind sources reach the receptor. In short, a composite descriptor represents the net of these paths as observed. The paths may differ in sign. Thus, the observed coefficient is a partial cancellation whose value depends on which path dominated in the fitting data.

Meteorological normalisation offers a partial response. Random-forest deweathering estimates meteorology's contribution to a concentration series and removes much of its variance, and \citep{grange2019using} use it to recover known intervention breakpoints, detecting the response at London Marylebone Road to an adjacent bus lane in 2001 and to the Congestion Charging Zone in February 2003, together with the effect of marine-fuel sulfur limits in Dover SO\textsubscript{2} series. What it removes, however, is the combined contribution of all paths, not a decomposition of them, and what it returns is a concentration series rather than an estimate of any other quantity. It is therefore a retrospective instrument: it detects the effect of a measure already taken, and does not extend to estimating the effect of one not yet taken, or to recovering the unobserved emissions and dispersion conditions that produced the observations.
 
\subsection{Relations among the abstracted variables}
\label{sec:mechanism-form}
The operations above concern what an abstraction discards. A distinct question concerns the relations that connect the surviving variables. High-level models assert relations among their coarse variables, and how those relations are established determines which of the three purposes the model can serve. The relations may be grouped into three classes according to how much of their structure comes from physics rather than data. In the first, the relation is derived from the governing equations. In the second, the structure is assumed on physical grounds and carried by the model, with observations supplying only what that structure leaves undetermined. In the third, the relation is fitted entirely, with its form chosen for approximation capacity rather than correspondence to any physical process. 

\paragraph{Relations derived from the fine-grained model} The form is not chosen but computed by operating on the governing equations or on the model that solves them. Galerkin projection substitutes a truncated basis into the equations and projects, ensuring the reduced system inherits the equation structure in reduced coordinates \citep{khamlich2023physics}. Adjoint sensitivity methods differentiate the chemical transport model, returning the response of a receptor to a perturbation at every source simultaneously \citep{zhang2025adjoint}, and brute-force apportionment obtains equivalent information by rerunning the model with a source removed. These are computationally demanding because they retain the fine-grained mechanism rather than replacing it. Therefore, they retain separability similar to the fine-grained model: each process is represented independently, e.g., advection is unaffected by a change to a chemical rate constant, a change in emissions alters the source term while leaving transport as it was.

\paragraph{Relations assumed on physical grounds, with parameters fitted} The model itself carries the system's physical structure, and observations supply only what that structure leaves undetermined. Separability follows from the representation. Because the model has components corresponding to distinct processes, one may be altered while the others remain unchanged. Source-receptor and transfer-coefficient matrices assume linear superposition of source contributions \citep{amann2011cost}, and chemical mass balance assumes conservative mixing of known source profiles \citep{hopke2016review}. The separation is therefore only as sound as the assumption that produced it, and where the assumed form misstates the physics, it is correspondingly misleading. Under linear superposition, the contribution attributed to a source and the change that would follow from removing it are the same quantity, whereas under non-linear chemistry they are distinct \citep{clappier2017source}.

\paragraph{Relations fitted without an assumed form} The relation is whatever minimises error against observations over a chosen function class, e.g., regression functions, polynomial and spline response surfaces, Gaussian-process and neural surrogates, or message passing over a graph \citep{beelen2013development, wang2020pm2}. The class is selected for approximation capacity rather than for correspondence to any physical process, and the fitted object consequently has no components corresponding to parts of the system. A regression coefficient is estimated jointly with every other coefficient against a single distribution and therefore has no interpretation apart from that estimation. A message-passing update describes information propagation during inference rather than influence propagation in the atmosphere, and fixing a node's value and evaluating the network forward does not correspond to altering any mechanism \citep{zevcevic2021relating}. Physics-informed approaches occupy an intermediate position, constraining the fitted function by the governing equations through a residual penalty \citep{raissi2019physics, shi2025phy}. This constraint improves physical fidelity by requiring the solution to satisfy the equations, but it remains a single function of its inputs, and the separability of the fine-grained representation is therefore not recovered.

Table~\ref{tab:review-synthesis} sets the models commonly used in urban air quality assessment against the scales at which they operate, the abstraction operations they embody, and the manner in which their relations are established. As shown in the table, the entries move from relations obtained by discretising or differentiating the governing equations, through relations whose form is assumed on physical grounds with the remainder calibrated or estimated, to relations fitted in their entirety against observations. Separability declines across this progression. A change acting on one part of the system has no counterpart in a non-separable model, since the model contains no component corresponding to the part that changed. A high-level relation supports a manipulation only where the component corresponding to the manipulated quantity may be altered while the remainder stands, and the same property governs inference in the reverse direction, since a quantity recovered from observation is interpretable only if it corresponds to something the model represents separately.

Reflecting on the three practical purposes of high-level models, tractability runs throughout, and the fitted families serve it most cheaply. However, abduction and intervention are served only when the abstraction leaves the model with components corresponding to the underlying physics: the abstraction operations determine whether such components survive coarsening, while the way the relations are established determines whether they can be altered one at a time. Alignment between a model and the observations it was fitted to guarantees neither. Only models that represent physical structure while fitting the remainder can serve all three purposes simultaneously: relations derived from the fine-grained model retain the cost the abstraction was introduced to avoid, and relations fitted without represented structure forgo what the upper two purposes require. Two questions follow, and they occupy the remainder of the paper. The first is what may serve as the represented physical structure, given that it must be specific enough to confer separability while leaving the model free to fit observations at acceptable cost. The second is what the abstraction must satisfy for that structure to remain valid at the coarse scale, and hence for observational and interventional queries to be answerable. 

\begin{table}[htbp]
\small
\centering
\caption{Models commonly used in urban air quality assessment}
\label{tab:review-synthesis}
\footnotesize
\begin{tabular}{@{}p{2.3cm}p{1.5cm}p{1.3cm}p{2.5cm}p{4.4cm}@{}}
\toprule
Model family & Spatial scale & Temporal scale & Abstraction operations & Establishment of the relation \\
\midrule
Street-canyon CFD &
1--10\,m & s--min &
None (Baseline as ground truth) &
Discretised governing equations \\
\addlinespace
Chemical transport model &
1--50\,km & hourly &
$\tau_{space}$, $\tau_{species}$, $\tau_{source}$ &
Discretised governing equations with a reduced chemical mechanism \\
\addlinespace
Gaussian plume / dispersion &
10\,m--1\,km & hourly &
$\tau_{space}$, $\tau_{met}$ &
Dispersion form assumed from turbulence theory; stability and source parameters supplied \\
\addlinespace
POD--Galerkin ROM &
Parent mesh & hourly &
$\tau_{space}$ &
Equations projected onto a basis extracted from simulation snapshots \\
\addlinespace
Adjoint sensitivity &
Parent grid & hourly &
$\tau_{source}$ &
Fine model differentiated about a base state \\
\addlinespace
Source--receptor matrix &
10--50\,km & annual &
$\tau_{source}$, $\tau_{met}$, $\tau_{time}$ &
Coefficients simulated by single-source perturbation; linear superposition assumed for combination \\
\addlinespace
Receptor apportionment (CMB, PMF) &
Site & daily--annual &
$\tau_{source}$, $\tau_{species}$ &
Conservative mixing assumed; contributions estimated from observed composition \\
\addlinespace
Land use regression &
25--100\,m & annual mean &
$\tau_{space}$, $\tau_{time}$, $\tau_{source}$, $\tau_{met}$ &
Coefficients fitted jointly against observations over an assumed predictor set \\
\addlinespace
Graph neural network &
Station or cell & hourly &
$\tau_{space}$, $\tau_{time}$ &
Adjacency assumed geometrically; update rule fitted \\
\addlinespace
Neural emulator &
Parent grid & hourly &
$\tau_{space}$, $\tau_{source}$ &
Input--output map fitted to simulated model responses \\
\bottomrule
\end{tabular}
\end{table}

Taken together, whether a coarse variable retains a physical referent and whether the relations among such variables may be altered independently are questions about the correspondence between a high-level model and the fine-grained system it stands for. Causal abstraction provides a standing formulation of that correspondence. Its central condition is that coarsening and intervening should yield the same result in either order, making sure that the response of the high-level model to a manipulation of an aggregate variable agrees with the coarsened response of the fine model to the corresponding physical change \citep{rubenstein2017causal, beckers2019abstracting}. The following section introduces the preliminaries of causal abstraction and states the requirements that abduction and intervention impose on high-level urban air quality models.

\section{Preliminaries}
Causal abstraction provides a theoretical foundation for when a coarse description of a system aligns with a fine-grained one. In urban air quality modelling, the fine-grained model is the discretised physics, the high-level model is the object on which reasoning can be conducted, and the abstraction map is the catalogued composition of operations in Section~\ref{sec:review}. The high-level model is not merely a low-cost substitute for the fine one. It is the level at which a question can be posed and an answer understood, since the variables it carries are the ones humans can reason over and act upon, e.g., a zone boundary, a speed limit, and a fleet composition, while a response expressed over a resolved atmospheric field would answer no question that had been asked. 

\paragraph{Notation and definitions} Uppercase letters ($X$) are used to denote variables and lowercase letters ($x$) to denote corresponding values. Bold uppercase ($\mathbf{X}$) and bold lowercase ($\mathbf{x}$) letters denote sets of variables and values, respectively. $D_X$ denotes the domain of $X$ and $D_{\mathbf{X}} = D_{X_1} \times \cdots \times D_{X_k}$ for the domain of $\mathbf{X} = \{X_1, \dots, X_k\}$, and $P(\mathbf{X} = \mathbf{x})$, often shortened to $P(\mathbf{x})$, for the probability of $\mathbf{X}$ taking values $\mathbf{x}$.
 
\paragraph{Structural causal models} The structural causal model (SCM) is adopted as it supports abduction and intervention. It assigns each variable its own mechanism, and a manipulation is represented by replacing one mechanism while the others remain. What the model omits is captured by exogenous variables, ensuring that influences acting on several quantities at once appear in the structure rather than disappearing into fitted coefficients. A SCM $\mathcal{M}$ of an urban atmosphere consists of endogenous variables $\mathbf{V}$ (e.g., concentrations at cells or receptors, emission strengths on links or sectors, aggregate activity variables, and measured meteorological conditions), exogenous variables $\mathbf{U}$ with distribution $P(\mathbf{U})$ (e.g., the pollutant background advected into the domain, and unrecorded disturbances to activity), and mechanisms $\mathcal{F}$, including functions $f_{V_i}$ for each $V_i \in \mathbf{V}$, mapping the endogenous parents $\mathbf{Pa}_{V_i}$ and exogenous parents $\mathbf{U}_{V_i}$ of $V_i$ to its value. $f_{V_i}$ states how the concentration at a cell arises from the emissions within it, the concentrations upwind, and the conditions under which the pollutant forms.

Each $\mathcal{M}$ induces a causal diagram $\mathcal{G}$, in which every $V_i \in \mathbf{V}$ is a vertex, a directed edge $V_j \to V_i$ is present for every $V_j \in \mathbf{Pa}_{V_i}$, and a dashed bidirected edge $V_j \dashleftarrow\dashrightarrow V_i$ is present for every pair whose exogenous parents are dependent. Bidirected edges are particularly consequential in this domain, because confounding factors can disturb several parts of the system at once. For instance, a traffic incident raises emissions on the affected corridor and on every route that traffic diverts onto, lifting concentrations at several receptors simultaneously. Two concentrations may therefore be associated without either causing the other, and conditioning on anything the model holds will not remove the association. Such pairs share dependent exogenous parents and are connected by bidirected edges.  Following \citep{xia2023neural}, our treatment is constrained to recursive SCMs, which induce acyclic diagrams. 
 
\paragraph{The causal hierarchy} Each SCM induces three layers of distributions, ordered by the inferential content they carry \citep{bareinboim2022pearl}. Layer $\mathcal{L}_1$ contains the observational distribution $P(\mathbf{V})$, which supports queries about the conditions under which the model was fitted, and, by conditioning, the recovery of unobserved quantities from measured ones. Layer $\mathcal{L}_2$ contains the interventional distributions $P(\mathbf{V}_{\mathbf{x}})$, obtained by replacing the mechanisms of the intervened variables with constants. Layer $\mathcal{L}_3$ contains counterfactual distributions, in which variables are evaluated in several worlds at once. Formally:
 
\begin{definition}[Layer 2 Valuation]
An SCM $\mathcal{M}$ induces layer $\mathcal{L}_2(\mathcal{M})$, a set of distributions over $\mathbf{V}$, each of the form $P(\mathbf{V}_{\mathbf{x}})$ such that
\begin{equation}
P^{\mathcal{M}}(\mathbf{v}_{\mathbf{x}}) = \int_{D_{\mathbf{U}}} \mathbb{1}\left[ \mathbf{V}_{\mathbf{x}}(\mathbf{u}) = \mathbf{v} \right] \, dP(\mathbf{u}),
\end{equation}
where $\mathbf{V}_{\mathbf{x}}(\mathbf{u})$ is evaluated under $\mathcal{F}_{\mathbf{x}} := \{ f_{V_j} : V_j \in \mathbf{V} \setminus \mathbf{X} \} \cup \{ f_X \leftarrow x : X \in \mathbf{X} \}$.
\end{definition}
 
The layers $\mathcal{L}_1$ and $\mathcal{L}_2$ correspond to the two purposes with which this paper is concerned. An \emph{observational query} asks for a quantity of $\mathcal{L}_1$. The conditional $P(\mathbf{V}_{\mathrm{unobs}} \mid \mathbf{v}_{\mathrm{obs}})$ recovers quantities that are not measured from those that are, for the purpose of abduction. For example, the emission strength on an unmonitored link. An \emph{interventional query} asks for a quantity of $\mathcal{L}_2$, the probability that the outcome variables $\mathbf{Y}$ take values $\mathbf{y}$ under the intervention setting $\mathbf{X}$ to $\mathbf{x}$, equivalently $P(\mathbf{Y} = \mathbf{y} \mid do(\mathbf{X} = \mathbf{x}))$. Inherently, SCM is capable of satisfying the separability requirement in  Section~\ref{sec:mechanism-form}, as a prerequisite of $\mathcal{L}_2$. 

The layers are strictly nested, and agreement at $\mathcal{L}_1$ places no constraint on $\mathcal{L}_2$. In other words, two models may induce identical observational distributions and differ arbitrarily in their interventional ones \citep{bareinboim2022pearl}. Whether an $\mathcal{L}_2$ quantity can be determined depends on the particular query. Given the available evidence and structure, some interventional queries are uniquely fixed while others are not, and this distinction depends on the query rather than on the model that answers it.
 
\paragraph{Query identification} Let $\mathbf{Z}$ denote the available evidence, comprising the observations as well as interventional information. A query $Q$ is identifiable from $\mathbf{Z}$ and $\mathcal{G}$ when every SCM consistent with both assigns it the same value. Identifiability is jointly determined by $\mathbf{Z}$ and $\mathcal{G}$, and it is not a matter of how much data is available. A query may remain non-identifiable, however comprehensive the observations, when several models may be equally consistent with $\mathbf{Z}$ while assigning $Q$ different values.

The neural parameterisation admits a procedure for deciding which case obtains \citep{xia2021causal}. Two models are fitted, both constrained to agree with $\mathbf{Z}$ and with $\mathcal{G}$, one minimising the query and one maximising it. Where the two converge, the evidence admits a single answer, the query is identified, and the common value is its estimate. Where they diverge, the interval between them is the range of answers the evidence permits, and the query is not identified. The same construction applies at $\mathcal{L}_1$. In other words, if an emission strength recovered from receptor readings is returned as a range, indicating that the monitoring network does not constrain the quantity rather than that the fitting has failed.

\paragraph{Neural causal models} The SCM fixes that each variable has its own mechanism, but does not specify what that mechanism is. A feasible solution is to implement neural parameterisation, i.e., an SCM whose mechanisms are neural networks, retaining a distinct mechanism per variable.
 
\begin{definition}[$\mathcal{G}$-Constrained Neural Causal Model]
\label{def:GNCM}
Given a causal diagram $\mathcal{G}$, a $\mathcal{G}$-constrained neural causal model $\widehat{\mathcal{M}}(\bm{\theta})$ over variables $\mathbf{V}$ is an SCM $\langle \widehat{\mathbf{U}}, \mathbf{V}, \widehat{\mathcal{F}}, \widehat{P}(\widehat{\mathbf{U}}) \rangle$ in which each $\hat{f}_{V_i} \in \widehat{\mathcal{F}}$ is a feedforward neural network parameterised by $\bm{\theta}_{V_i}$, mapping the endogenous parents of $V_i$ in $\mathcal{G}$ together with the exogenous variables assigned to $V_i$ by the bidirected structure of $\mathcal{G}$, and each $\widehat{U} \in \widehat{\mathbf{U}}$ is distributed uniformly on $[0,1]$.
\end{definition}

Leveraging NCM as the high-level model $\mathcal{M}_H$ for a fine-grained atmospheric model $\mathcal{M}_L$. The micro-variables $\mathbf{V}_L$ are the resolved fields, and the macro-variables $\mathbf{V}_H$ are the coarse quantities that survive abstraction. An abstraction map $\tau$ assigns a coarse state to each fine state: $D_{\mathbf{V}_L} \to D_{\mathbf{V}_H}$, and a second map $\omega$ relates the interventions available between the high and low levels.

\begin{definition}[Interventional Consistency]
\label{def:consistency}
The pair $(\mathcal{M}_H, \tau)$ is a causal abstraction of $\mathcal{M}_L$ with respect to an intervention set $\mathcal{I}$ if, for every $i \in \mathcal{I}$,
\begin{equation}
\tau \big( \mathcal{L}_2^{\mathcal{M}_L}(i) \big) = \mathcal{L}_2^{\mathcal{M}_H}\big( \omega(i) \big),
\end{equation}
that is, abstracting the fine model's response to an intervention yields the high-level model's response to the corresponding high-level intervention.
\end{definition}

The condition is a commuting requirement: coarsening and intervening must yield the same result in either order. It cannot, however, be verified in practice as interventional distributions of $\mathcal{M}_L$ are unavailable. What follows shows that it can nonetheless be secured by conditions that can be checked without implementing intervention.

\begin{proposition}[Interventional Warrant without Interventional Evidence]
\label{prop:warrant}
Let $\omega$ be well defined over $\mathcal{I}$, let $\mathcal{G}_H$ be a causal diagram of the coarse system, let $\mathcal{M}_H$ be $\mathcal{L}_1$-consistent with the available evidence $\mathbf{Z}$, and let $Q \in \mathcal{L}_2$ be identifiable from $\mathbf{Z}$ and $\mathcal{G}_H$. Then the value $\mathcal{M}_H$ assigns to $Q$ is the value assigned by every SCM consistent with $\mathbf{Z}$ and $\mathcal{G}_H$, and hence by the true high-level model.
\end{proposition}

In short, where $\mathcal{M}_H$ is a $\mathcal{G}_H$-constrained NCM fitted to $\mathbf{Z}$, the identifiability of a query is decidable from $\mathcal{G}_H$, and an identified query can be answered correctly provided the first two conditions hold. Neither is supplied by the fitting. The first requires that a coarse intervention determine its fine-level counterpart: many configurations of the resolved field realise the same coarse value, and where these disagree on the outcome, there is no single fine-level answer for the coarse model to match. The second requires that $\mathcal{G}_H$ record the mechanisms of the coarse system. Identifiability is therefore necessary but not sufficient, and a query identified under a diagram that misrepresents the system, or expressed over an aggregate whose realisations disagree, is answered with unwarranted confidence.
 
In practice, the outputs of $\mathcal{G}_H$-constrained NCM and the observations $\mathbf{Z}$ will not agree exactly. The chemistry is non-linear, coarsening discards variation, and abstraction error between $\mathcal{M}_H$ and $\mathcal{M}_L$ is unavoidable. The criterion is applied in its approximate form, with a discrepancy $\varepsilon$ measuring how far the high-level model's response departs from the coarsened response of the fine one \citep{beckers2020approximate}. 

\section{Constructing High-Level Models for Urban Air Quality}
The three purposes for which high-level models are constructed impose increasingly strong requirements. Tractability requires only a computable $\tau$ and an $\mathcal{M}_H$ cheap to evaluate relative to $\mathcal{M}_L$. Abduction requires $\mathcal{L}_1$-consistency, meaning the observational distribution of $\mathcal{M}_H$ agrees with that of $\mathcal{M}_L$ after coarsening by $\tau$, and it also requires the inferred quantity to be constrained by what is observed. The second condition does not follow from the first. A model may reproduce the observations while the emission strength on an unmonitored link remains underdetermined, because sparse receptors and multiple contributing sources leave many emission fields consistent with the same readings. Intervention imposes additional conditions. Each variable of $\mathcal{M}_H$ must be produced by an individual mechanism. For example, a low-emission zone assigns a new value to the emissions of the affected links while dispersion and chemistry remain unaltered. The diagram $\mathcal{G}_H$ must record these mechanisms rather than the associations present in the observations. An edge from an upwind cell to a receptor asserts that the concentration in the upwind cell is an argument of the mechanism producing the concentration at the receptor. And the query must be identifiable from $\mathcal{G}_H$ and the evidence, since distinct models may reproduce the observations to the same degree while assigning different values to a given query.

The neural causal model meets all these requirements, and this section outlines its construction. Figure~\ref{fig:causal} sets out the relation that the construction must establish. First, the abstraction map $\tau$ is fixed, determining the macro-variables and the scales at which they are defined. Section~\ref{sec:admissibility} establishes the conditions a choice of $\tau$ must satisfy for the resulting variables to admit a causal structure. The mechanisms relating those variables are then represented in $\mathcal{G}_H$, as Section~\ref{sec:construction} sets out. Fitting $\mathcal{M}_H$ to the observations secures $\mathcal{L}_1$-consistency. Interventional consistency in $\mathcal{L}_2$ cannot be established directly, since the interventional distributions of $\mathcal{M}_L$ are unavailable. Identifiability takes its place, and Section~\ref{sec:queries} gives the procedure that decides it.

\input{Figures/Causal_abstraction}

\subsection{Admissibility of the abstraction}
\label{sec:admissibility}
The neural causal model carries prerequisites. It assumes an acyclic diagram $\mathcal{G}_H$ whose variables are Markov with respect to it, and neither property is guaranteed by default. Both depend on the abstraction map $\tau$, and specifically on the cell size $L$ and the reporting interval $\Delta t$ at which the coarse variables are defined. This section establishes the conditions.

The reporting interval $\Delta t$ is the temporal component of the abstraction operation $\tau$. A high-level model reports a single value per cell per interval, and that value averages everything the atmosphere did within it. Therefore, the choice of $\Delta t$ determines how much of the dynamics is absorbed into a node rather than represented between nodes. The interval is bounded from both directions: too long and the coarse variables cease to carry a causal structure, too short and the abstraction is no longer licensed to collapse the processes it collapses.

Specifically, two failures follow directly from $U \cdot \Delta t > L$, where $U$ is the effective advective wind at model level. Influence then crosses more than one cell within a single interval, and a diagram whose edges connect only adjacent cells is misspecified, forcing local edges to carry influence that originated further upwind. Markovianity fails for a related but distinct reason: a parcel crossing several cells within $\Delta t$ accumulates emissions and undergoes chemistry en route, and neither is recorded in any variable at time $t$, and therefore, the state at $t$ does not determine the state at $t + \Delta t$.

The acyclicity assumption fails under a separate condition. Under a steady wind, the unrolled diagram remains acyclic at any advective range, with influence flowing downwind alone. Cycles arise where the wind direction varies within the interval, or across the ensemble of intervals over which reporting is aggregated, since one cell may then influence another and be influenced by it within what is recorded as a single step. A long interval makes such variation more likely, but the operative condition is the steadiness of the wind over $\Delta t$ rather than the distance influence travels within it.

On the other end, the abstraction collapses sub-cell processes into single nodes, and the node value stands for the state those processes have reached. This is licensed only when they have relaxed within the interval. A fast process at its stationary state admits a static, acyclic representation, whereas one still evolving does not. Street-canyon recirculation and the photostationary state of the NO--NO\textsubscript{2}--O\textsubscript{3} system are the cases at issue here, and where $\Delta t$ is shorter than their relaxation times, the node carries no well-defined value, and the mechanism producing it is not a function of the coarse variables.

\begin{proposition}[Admissible High-Level Abstraction]
\label{prop:window}
Let $L$ be the cell size, $U$ the effective advective wind speed at model level, $\Delta t$ the reporting interval, and $\tau_{\mathrm{equil}}$ the longest relaxation time among the sub-cell processes collapsed into single nodes. A high-level diagram $\mathcal{G}_H$ over the field admits an acyclic and Markovian causal structure only if
\begin{equation}
\tau_{\mathrm{equil}} \; \ll \; \Delta t \; \lesssim \; L / U,
\label{eq:window}
\end{equation}
with the wind direction approximately steady over $\Delta t$.
\end{proposition}
  
The bounds of Proposition~\ref{prop:window} are evaluated below under conditions representative of urban modelling. The lower bound has two contributors. Street-canyon ventilation proceeds through roof-level exchange at a velocity that is a small fraction of the reference wind. The exchange velocity is reported in values of order 1\% of the free-stream wind for a dense canyon, rising to several percent for sparser arrays \citep{solazzo2007transfer}. Taking $u_{\mathrm{exch}} \approx 0.03U$ and a canyon of height $H = 20$\,m at $U = 3$\,m$\cdot$\,s$^{-1}$ gives a flushing time $\tau_{\mathrm{flush}} \approx H / u_{\mathrm{exch}} \approx 220$\,s, of order three to four minutes, and longer for deeper canyons or under stable stratification. The photostationary state of the NO--NO\textsubscript{2}--O\textsubscript{3} system relaxes on $\tau_{\mathrm{PSS}} \approx 1/j(\mathrm{NO_2})$, which for a clear-sky daytime value of $j(\mathrm{NO_2}) \approx 8 \times 10^{-3}$\,s$^{-1}$ is approximately two minutes \citep{madronich1983measurement}. Taking $\tau_{\mathrm{equil}} \approx 4$\,min and reading $\ll$ as a factor of five gives $\Delta t \gtrsim 20$\,min. Denser canyons flush more slowly and would raise this bound, and the figure adopted here is the one most favourable to a short reporting interval.

Given the lower bound, whether any admissible abstraction exists depends on the upper bound. Table~\ref{tab:window} illustrates the advective crossing time $L/U$ in minutes across the cell sizes and wind speeds. The speed $U$ is the mass-weighted mean horizontal wind over the vertical extent that the coarse variable represents, evaluated from the fields of $\mathcal{M}_L$, and the range of 1 to 10\,m$\cdot$\,s$^{-1}$ spans the typical conditions urban and regional modelling encounters. At one-kilometre resolution, the upper bound is 5.6\,min at $U = 3$\,m$\cdot$\,s$^{-1}$, against a lower bound of 20\,min, and the two cross only below $U \approx 0.8$\,m$\cdot$\,s$^{-1}$, which lies beneath the threshold at which wind is conventionally treated as calm. No reporting interval, whatever, yields a one-kilometre diagram $\mathcal{G}_H$. The tension does not stem from any demand for street-level detail. Chemical transport models are ordinarily run at half a kilometre to two kilometres over a city domain, and at those sizes air crosses in three to eleven minutes, far short of the twenty required.

\begin{table}[htbp]
\centering
\caption{Advective crossing time $L/U$ in minutes as a function of cell size $L$ and effective wind speed $U$ in m$\cdot$\,s$^{-1}$}
\label{tab:window}
\small
\begin{tabular}{@{}*{6}{>{\centering\arraybackslash}p{1.5cm}}@{}}
\toprule
$L$ & $U = 1$ & $U = 2$ & $U = 3$ & $U = 5$ & $U = 10$ \\
\midrule
1\,km & 16.7 & 8.3 & 5.6 & 3.3 & 1.7 \\
2\,km & \textbf{33.3} & 16.7 & 11.1 & 6.7 & 3.3 \\
5\,km & \underline{\textbf{83.3}} & \textbf{41.7} & \textbf{27.8} & 16.7 & 8.3 \\
10\,km & \underline{\textbf{166.7}} & \underline{\textbf{83.3}} & \textbf{55.6} & \textbf{33.3} & 16.7 \\
20\,km & \underline{\textbf{333.3}} & \underline{\textbf{166.7}} & \underline{\textbf{111.1}} & \underline{\textbf{66.7}} & \textbf{33.3} \\
\bottomrule
\end{tabular}

\vspace{4pt}
\begin{minipage}{\textwidth}
\footnotesize Note: Entries in bold exceed the lower bound $\Delta t \gtrsim 20$\,min and therefore admit a non-empty window, within which $\Delta t$ must lie between 20\,min and the tabulated value. Underlined entries exceed 60\,min and therefore admit hourly reporting.
\end{minipage}
\end{table}

A different choice of macro-variable avoids the issue. An Eulerian cell asserts that the pollutant reaches its neighbour within one interval, and that assertion fails once $U \cdot \Delta t$ exceeds $L$. A source--receptor formulation makes no such assertion. Its macro-variables are the emissions of identified sources and the concentrations at identified receptors, with the relation between them carried by a footprint stating how much each source contributes to each receptor under the prevailing conditions. The distance the air travels and the time it takes are absorbed into that relation rather than claimed by the structure of the diagram, and the upper bound accordingly has no application. Acyclicity holds for the same reason: sources emit, and receptors receive, and no edge reverses however the wind turns. The lower bound is unchanged for the source--receptor formulation. A receptor value still stands for processes that must have settled within the interval. And where transit from a source exceeds one interval, the footprint depends on conditions prevailing earlier. The mechanism must carry lagged conditions among its arguments, the required extent being set by the longest transit time in the domain. The cost of the source--receptor formulation is a narrowing of what the model delivers: concentrations are available only at the receptors and nowhere else. 
 
\subsection{Construction}
\label{sec:construction}

Proposition~\ref{prop:window} constrains the macro-variables and the transport edges among them, leaving the remaining structure and the mechanisms unsettled. This section sets out the construction of $\mathcal{M}_H$: how the generation of emissions, the persistence and transport of pollutants, and chemical transformation are represented in $\mathcal{G}_H$ and parameterised in $\mathcal{M}_H$, and how a candidate model $\mathcal{\hat{M}}_H$ is fitted and tested.

The endogenous variables are, for each cell $i$ and interval $t$: the concentration $C_{i,t}^{s}$ of each species $s$, the emission $E_{i,t}^{s}$ of each species from sources within the cell, the activity variables $A_{i,t}$ that determine emission, such as traffic volume by vehicle class and heating demand, and the meteorological conditions $W_{i,t}$, comprising wind, mixing height, temperature and radiation.

Emissions are typically unobservable. Therefore, the emission variable is carried as a latent endogenous variable with its own mechanism,
\begin{equation}
E^{s}_{i,t} = \hat{f}^{s}_{E}\left( A_{i,t},\, W_{i,t},\, \widehat{U}^{E}_{i,t} \right),
\label{eq:emis}
\end{equation}
taking activity as its principal argument, the meteorological conditions that modulate emission, and exogenous variables $\widehat{U}^{E}_{i,t}$ carrying what the representation omits. 

A cell retains part of its own previous concentration. What remains after deposition, after dilution as the mixing height changes, and after outward advection constitutes a persistence term, and $C_{i,t-1}^{s}$ is accordingly a parent of $C_{i,t}^{s}$. The term depends on conditions through deposition velocity and mixing height, and its magnitude sets the rate at which a cell forgets its own state. The remaining transport parents $\mathbf{Pa}^{\mathrm{ch}}_s$ are the concentrations $C_{j,t-1}^{s}$ of every cell $j$ within advective range of $i$, together with $W_{i,t}$ and $W_{j,t-1}$. An edge from $j$ to $i$ asserts only that transport between them is possible. Whether transport occurs, and with what strength, is not represented in $\mathcal{G}_H$ but learned from the data by the mechanism, conditioned on $W$. The diagram therefore admits more parents than convey influence at any given time, and an edge that is inactive throughout most of the record is not thereby misspecified.

The parents of $C_{i,t}^{s}$ also include the concentrations $C_{i,t-1}^{s'}$ of the other species in the same cell, and this edge set is supplied by domain knowledge. A chemical mechanism states which species participate in reactions that produce and consume $s$. Such knowledge is type-level: the reaction holds wherever the species are present, and the same template instantiates at every cell and interval. The mechanism does not state whether an edge is active at a given operating point, with what strength, or with what sign. The sensitivity of ozone to NO\textsubscript{x} reverses across the boundary between the NO\textsubscript{x}-limited and VOC-limited regimes, which suggests that the sign is a property of the operating point rather than of the reaction. As with the transport parents, the edges record which species may depend on $s$, while the magnitude and sign of that effect are estimated from data.

Each mechanism is a feed-forward network, and the mechanism producing a concentration is decomposed according to the processes contributing to it. The concentration variable combines mass carried over from the previous interval, released within the cell, and advected from upwind, together with the net production or loss of the species by chemical reaction,
\begin{align}
C_{i,t}^{s} \;=\;& \hat{g}^{\,s}_{P}\big(C_{i,t-1}^{s},\, W_{i,t}\big)
\;+\; \hat{g}^{\,s}_{E}\big(E_{i,t}^{s},\, W_{i,t}\big)
\;+\; \sum_{j \in \mathbf{Pa}^{\mathrm{tr}}_i} \hat{g}^{\,s}_{\mathrm{tr}}\big(C_{j,t-1}^{s},\, W_{i,t},\, W_{j,t-1}\big) \nonumber \\
&+\; \hat{g}^{\,s}_{\mathrm{ch}}\big(\{C_{i,t-1}^{s'}\}_{s' \in \mathbf{Pa}^{\mathrm{ch}}_s},\, C_{i,t-1}^{s},\, W_{i,t}\big)
\;+\; \widehat{U}^{C}_{i,t},
\label{eq:conc}
\end{align}
in which $\hat{g}^{\,s}_{P}$ carries what remains of the cell's own previous concentration after deposition and dilution, $\hat{g}^{\,s}_{E}$ converts the emission released within the cell into the concentration increment it produces under the given mixing conditions, $\hat{g}^{\,s}_{\mathrm{tr}}$ indicates the contribution advected from each cell $j$ within range, and $\hat{g}^{\,s}_{\mathrm{ch}}$ takes the concentrations of the species in $\mathbf{Pa}^{\mathrm{ch}}_s$, those which the mechanism identifies as participating in the production or loss of $s$, together with the conditions governing the reaction rates. The exogenous variable $\widehat{U}^{C}_{i,t}$ is assigned by the bidirected structure of $\mathcal{G}_H$. 

With $\mathcal{G}_H$ specified and its mechanisms parameterised, what remains is to secure $\mathcal{L}_1$-consistency: the network weights are estimated from the observations $\mathbf{Z}$ by minimising $\mathcal{J}$, the negative log-likelihood of the observed quantities under $\widehat{\mathcal{M}}_H$. Consistency at $\mathcal{L}_1$ does not ensure the correctness of $\mathcal{G}_H$. An exogenous variable acting on two or more endogenous variables (a.k.a. a confounder) is readily omitted from the diagram because exogenous variables are unmeasured, and its influence is absorbed into the mechanisms of the variables on which it affects. The fitted model $\widehat{\mathcal{M}}_H$ fits the observations well, and the omission is not apparent from the fit. However, the omission biases the estimated mechanisms, potentially crediting endogenous variables with effects that belong to the exogenous one.

Residual analysis can detect such an omission. Where $\mathcal{G}_H$ is complete, what remains of a variable after its mechanism has accounted for it is exogenous noise, and this noise is independent across variables. A confounder breaks that independence because its influence remains in the residuals of every variable on which it affects, and those residuals are therefore correlated. Correlation between the residuals of two variables that $\mathcal{G}_H$ treats as independently disturbed indicates a confounder the diagram has not recorded. Algorithm~\ref{alg:fit} performs the fitting to obtain $\widehat{\mathcal{M}}_H(\bm{\theta}^{\ast})$ and looks for missing confounders.

\begin{algorithm}[h]
\caption{Estimation of $\mathcal{M}_H$ and diagnosis of missing confounders}
\label{alg:fit}
\begin{algorithmic}[1]
\Require Observations $\mathbf{Z}$; diagram $\mathcal{G}_H$
\State Partition $\mathbf{Z}$ into $\mathbf{Z}_{\mathrm{tr}}$ and $\mathbf{Z}_{\mathrm{te}}$ by contiguous blocks in time
\State Construct $\widehat{\mathcal{M}}_H(\bm{\theta})$ with one network per variable over the parent sets of $\mathcal{G}_H$
\State $\bm{\theta}^{\ast} \gets \arg\min_{\bm{\theta}} \; \mathcal{J}\big(\widehat{\mathcal{M}}_H(\bm{\theta}), \mathbf{Z}_{\mathrm{tr}}\big)$ \Comment{$\mathcal{J}$ the negative log-likelihood}
\State Evaluate $\mathcal{J}$ on $\mathbf{Z}_{\mathrm{te}}$
\State $r_V \gets$ residual of $V$ on $\mathbf{Z}_{\mathrm{te}}$ under $\widehat{\mathcal{M}}_H(\bm{\theta}^{\ast})$, for each $V \in \mathbf{V}$
\For{each pair $(V, V')$ without a bidirected edge in $\mathcal{G}_H$}
  \State Report $(V, V')$ where the correlation of $r_V$ and $r_{V'}$ exceeds the spread obtained under permutation
\EndFor
\State \Return $\widehat{\mathcal{M}}_H(\bm{\theta}^{\ast})$, $\mathcal{J}$ on each partition, and the reported pairs
\end{algorithmic}
\end{algorithm}
 
A reported pair $(V, V')$ calls for reconsideration of the diagram but does not determine what should change in $\mathcal{G}_H$. Correlated residuals indicate a shared influence, and the response may be to add a bidirected edge between $V$ and $V'$, to promote the influence to an endogenous variable where it is measured, or to revise the mechanisms. 
 
\subsection{Observational and interventional query}
\label{sec:queries}
 
The fitted model $\widehat{\mathcal{M}}_H(\bm{\theta}^{\ast})$ is capable of answering observational and interventional queries at $\mathcal{L}_1$ and $\mathcal{L}_2$. The queries of both layers are illustrated with the emission variables $E_{i,t}^{s}$. The query applies unchanged to any variable in $\mathcal{G}_H$. At $\mathcal{L}_1$ the model answers:
\begin{equation}
P\big( E_{i,t}^{s} \mid \{C_{r,t}^{s}\}_{r \in \mathcal{R}},\; A,\; W \big),
\label{eq:abduction}
\end{equation}
the emission at a cell given the concentrations measured at the receptor cells $r \in \mathcal{R}$, together with the activity and meteorological variables. 

A model $\widehat{\mathcal{M}}_H(\bm{\theta}^{\ast})$ that reproduces the observations $\mathbf{Z}$ does not always answer a query definitely. The emission case illustrates this. A receptor measures the total concentration, to which every upwind cell within range has contributed. Under a steady wind, the same cells contribute in the same proportions on every occasion, and the measurement constrains only their combined contribution: a large emission from one cell and a small one from another produce the same reading as the reverse, and the receptor cannot distinguish them. When the wind veers, the same cells contribute in different proportions, and a second measurement constrains a different combination of the same emissions. The posterior for any one emission is therefore wide where the wind is persistent, and narrower where the observations cover a range of wind directions.

At $\mathcal{L}_2$, an intervention acting on cells $\mathcal{S}$ over an interval $\mathcal{T}$ defines
\begin{equation}
Q \;=\; \mathbb{E}\big[\, C_{r,t}^{s} \;\big|\; \mathrm{do}\big( \{E_{i,\tau}^{s}\}_{i \in \mathcal{S}, \tau \in \mathcal{T}} = \mathbf{e} \big),\; W = w \,\big],
\label{eq:query}
\end{equation}
evaluated by replacing $\hat{f}^{\,s}_{E}$ at each $(i,\tau) \in \mathcal{S} \times \mathcal{T}$ with the constant $\mathbf{e}$, the terms $\hat{g}^{\,s}_{P}$, $\hat{g}^{\,s}_{\mathrm{tr}}$ and $\hat{g}^{\,s}_{\mathrm{ch}}$ standing unaltered. The difference between this value and the one obtained under the unmodified emissions is the change at receptor $r$ that the intervention would produce.
 
Both queries are evaluated using the parameters $\bm{\theta}^{\ast}$ that Algorithm~\ref{alg:fit} returns, and both yield a value for any $\bm{\theta}^{\ast}$ obtained. A query is identifiable only when every parameterisation consistent with $\mathbf{Z}$ and $\mathcal{G}_H$ agrees on the same value. This condition is decided by searching the admissible parameterisations for the two that minimise and maximise the query while remaining consistent with $\mathbf{Z}$, following the procedure of \citep{xia2021causal}, as adapted in Algorithm~\ref{alg:id}.  
\begin{algorithm}[htbp]
\caption{Identification of a query under $\mathcal{G}_H$ and $\mathbf{Z}$}
\label{alg:id}
\begin{algorithmic}[1]
\Require Query functional $Q(\cdot)$; observations $\mathbf{Z}$; diagram $\mathcal{G}_H$; tolerance $\delta$
\Ensure $Q$ evaluated under $\widehat{\mathcal{M}}_H$ if identifiable, \textsc{fail} otherwise
\State $\widehat{\mathcal{M}}_H \gets$ NCM over $\mathbf{V}$ constrained by $\mathcal{G}_H$
\State $\mathcal{J}^{\ast} \gets \min_{\bm{\theta}} \; \mathcal{J}\big(\widehat{\mathcal{M}}_H(\bm{\theta}), \mathbf{Z}\big)$
\State $\bm{\theta}^{\min} \gets \arg\min_{\bm{\theta}} \; Q\big(\widehat{\mathcal{M}}_H(\bm{\theta})\big)$ \; s.t. \; $\mathcal{J}\big(\widehat{\mathcal{M}}_H(\bm{\theta}), \mathbf{Z}\big) \leq \mathcal{J}^{\ast} + \epsilon$
\State $\bm{\theta}^{\max} \gets \arg\max_{\bm{\theta}} \; Q\big(\widehat{\mathcal{M}}_H(\bm{\theta})\big)$ \; s.t. \; $\mathcal{J}\big(\widehat{\mathcal{M}}_H(\bm{\theta}), \mathbf{Z}\big) \leq \mathcal{J}^{\ast} + \epsilon$
\If{$Q\big(\widehat{\mathcal{M}}_H(\bm{\theta}^{\min})\big) \neq Q\big(\widehat{\mathcal{M}}_H(\bm{\theta}^{\max})\big)$}
  \State \Return \textsc{fail}
\Else
  \State \Return $Q\big(\widehat{\mathcal{M}}_H(\bm{\theta}^{\min})\big)$
\EndIf
\end{algorithmic}
\end{algorithm}
 
Both extrema are obtained by optimising $Q(\bm{\theta}) \pm \lambda\, \mathcal{J}\big(\widehat{\mathcal{M}}_H(\bm{\theta}), \mathbf{Z}\big)$, with $\lambda$ set large enough that the fit is not sacrificed to move the query. With finite observations, the model cannot match $\mathbf{Z}$ exactly, and lines 3 and 4 admit any $\bm{\theta}$ whose loss is within $\epsilon$ of the best attainable. The result is then a gap between the maximum and the minimum. A larger $\epsilon$ admits more models and widens the gap. As long as the gap between the maximum and the minimum closes as $\epsilon$ is reduced, the query is identifiable. If it does not close, models that fit the observations equally well still disagree about the query, and no further observation of the same kind will settle it. 

Where a query is not identifiable, the obstruction lies in the bidirected structure of $\mathcal{G}_H$. For a single intervened variable, the effect is identifiable exactly when no bidirected path connects that variable to any of its children \citep{shpitser2008complete}. In this sense, the two algorithms are complementary: Algorithm~\ref{alg:fit} establishes which confounding $\mathcal{G}_H$ must carry, and Algorithm~\ref{alg:id} establishes which queries the resulting diagram can answer. 

\section{Controlled Synthetic Experiment}
This section demonstrates the construction of a high-level air quality model on a synthetic system. The purpose is not to reproduce the empirical complexity of a particular city but to provide a setting in which the fine-grained mechanisms, the confounding structure, the abstraction map and the response to intervention are known by construction. Because they are known, every answer the high-level model returns can be checked against the system that produced the data, and an omission in the causal diagram can be distinguished from an error in estimating its mechanisms, which observation alone cannot separate. The claims that follow accordingly concern the conditions under which a high-level model supports causal reasoning, and are not empirical findings about any city's air.

\subsection{The fine-grained model $\mathcal{M}_L$}
\label{sec:low-level}
The synthetic system is a small city laid out along a prevailing wind, illustrated in Figure~\ref{fig:layout}. Four zones sit in sequence. An industrial district occupies the upwind edge, with heavy power plants running continuously. A traffic corridor follows, carrying the arterial route into the city and its two daily peaks. Beyond it lies a residential district, where emissions rise with heating demand on cold days. A suburban fringe sits at the downwind end, receiving what the city has produced and emitting little of its own. Each name records the dominant source rather than the only one, and every zone carries activity of the other kinds at lower intensity. Over the simulated period, construction projects begin and end in the industrial and traffic zones. While a project runs, it emits through non-road machinery during working hours and raises dust in the meantime. 

\input{Figures/Conceptual_system}

$\mathcal{M}_L$ is a discrete-time stochastic SCM over four zones. Each zone carries concentrations of NO\textsubscript{x}, PM\textsubscript{2.5} and SO\textsubscript{2} at hourly resolution, driven by emissions from four sources: traffic, industrial, residential and construction. Weather drives emissions and pollutant transport. Wind speed and mixing height follow first-order autoregressive processes that revert to climatological means of $\mu_{\text{wind}} = 4$\,m$\cdot$s$^{-1}$ and $\mu_{\text{mixing}} = 800$\,m. Temperature follows the same process, with a diurnal target ranging 6\,$^{\circ}$C on either side of a 15\,$^{\circ}$C mean. Wind speed and mixing height reach the concentrations through the dispersion index, $D_t = (\text{wind}_t \cdot \text{mixing}_t) / (\mu_{\text{wind}} \cdot \mu_{\text{mixing}})$, normalised to ensure $D_t \approx 1$ under mean conditions.

The emission of source $s$ in zone $z$ at hour $t$ is $E_{t,z,s} = \kappa_{z,s}\,\phi_s(t) + \varepsilon_{t,z,s}$, where $\kappa_{z,s}$ is a scale constant for that source in that zone, $\phi_s(t)$ is a temporal shape particular to the source, and $\varepsilon$ is Gaussian noise. Table~\ref{tab:scales} gives $\kappa_{z,s}$. The shapes $\phi_s(t)$ differ by source: traffic emissions carry two Gaussian peaks at 08:00 and 18:00, industrial emissions are constant, and residential emissions follow a baseline that rises linearly with the shortfall below 16\,$^{\circ}$C, at 0.12 per degree against a base of 0.3. A source's emissions are distributed across the three pollutants by fixed weights. Traffic contributes to NO\textsubscript{x} at 1.00 and to PM\textsubscript{2.5} at 0.25. Industrial sources contribute 0.60, 0.12 and 1.00 to NO\textsubscript{x}, PM\textsubscript{2.5} and SO\textsubscript{2}, respectively. Residential sources contribute 0.10, 0.40 and 0.30. 

\begin{table}[htbp]
\centering
\caption{Scale constant $\kappa_{z,s}$ for each zone and source}
\label{tab:scales}
\small
\begin{tabular}{@{}l*{3}{>{\centering\arraybackslash}p{2.2cm}}@{}}
\toprule
Zone & Traffic & Industrial & Residential \\
\midrule
Industrial  & 2.0  & \textbf{30.0} & 1.0 \\
Traffic     & \textbf{20.0} & 2.0  & 3.0 \\
Residential & 6.0  & 1.0  & \textbf{14.0} \\
Suburban    & 3.0  & 0.5  & 10.0 \\
\bottomrule
\end{tabular}
\end{table}

Not all particulate matter is emitted. A second portion of PM\textsubscript{2.5} forms in the air from NO\textsubscript{x} and SO\textsubscript{2} as precursor gases at the rate $\sqrt{C_{t,z,\mathrm{NO_x}} \cdot C_{t,z,\mathrm{SO_2}}} \cdot \max(0, (T_t - 15)/10)$. The temperature factor stands in for the role of warmth and sunlight in driving this chemistry: it is zero below 15\,$^{\circ}$C and rises linearly above, reaching unity at 25\,$^{\circ}$C. 

In addition to the conventional sources, $\mathcal{M}_L$ carries construction emissions, driven by an unobserved exogenous variable. A project starts in any hour with probability 0.005 and runs for 5 to 30 days. Its magnitude is fixed for the duration, gated to working hours by a smooth plateau between 07:00 and 18:00. It raises NO\textsubscript{x} through emissions from industrial and traffic zones, and raises PM\textsubscript{2.5} through those emissions and through fugitive dust added directly to the concentration. Because it affects two pollutants at once and is absent from the observations, the construction project is the confounder in the system.

A zone's concentration in a given hour comes from three sources. Most of what was present the hour before remains, decaying at a rate that differs by pollutant, and NO\textsubscript{x} and SO\textsubscript{2} clear faster than PM\textsubscript{2.5}. A share of the upwind zone's concentration arrives, proportional to wind speed, and the industrial zone at the head of the chain receives nothing. Add to that what the zone has emitted, divided by the dispersion index. The same emission therefore produces a higher concentration when the air is still, and the mixing layer is shallow. The secondary formation term is also added to PM\textsubscript{2.5}, followed by the construction dust. 

\subsection{The abstraction $\tau$ and the causal diagram $\mathcal{G}_H$ for $\mathcal{M}_H$}
\label{sec:GH}

The abstraction coarsens the four zones into two regions. Industrial and traffic form an urban region, and residential and suburban form a suburban one, so that $\mathcal{M}_H$ carries six concentration variables, one per region and pollutant, at hourly resolution.

Concentrations are aggregated from zones to regions by a population-weighted power mean,
\begin{equation}
\tilde{C}_{t,r,p} = \Big(\sum_{z \in \mathcal{Z}(r)} \bar{\omega}_z\, C_{t,z,p}^{\,\gamma}\Big)^{1/\gamma},
\qquad \bar{\omega}_z = \frac{\omega_z}{\sum_{z' \in \mathcal{Z}(r)} \omega_{z'}},
\label{eq:tau-conc}
\end{equation}
with $\gamma = 4$ and populations of 5{,}000, 20{,}000, 50{,}000 and 30{,}000 for the industrial, traffic, residential and suburban zones. The power biases the aggregate toward the more polluted member zone rather than averaging across the pair, as an exposure-weighted indicator requires.

$\mathcal{M}_H$ observes the weather, the region-level concentrations, the activity of the three conventional sources, and the hour through the harmonics of the diurnal cycle. The causal diagram $\mathcal{G}_H$ is shown in Figure~\ref{fig:GH}. Advection reaches only the suburban region, with the urban region heading the chain, and each suburban node takes as its advection parent the same pollutant in the urban region at $t-1$. The confounder $U_C$ has two children, urban emissions and urban PM\textsubscript{2.5}, while the emissions are themselves a parent of PM\textsubscript{2.5}. With the bidirected edge established according to Definition~\ref{def:GNCM}, this becomes the obstruction to interventional identification in $\mathcal{G}_H$, as Section~\ref{sec:queries} confirms.

\input{Figures/Causal_diagram}

Each variable is assigned a feedforward network over its parents in $\mathcal{G}_H$, according to Section~\ref{sec:construction}. The concentration is additive across the processes that contribute to it, with sub-networks representing persistence, advection, emission, and secondary chemistry, respectively. Therefore, an intervention can be imposed on one of them, such as the emission, which is what the interventional queries of Section~\ref{sec:queries} require.

The emission components take the inputs their sources respond to. The traffic component takes both harmonics of the diurnal cycle, since we assume two peaks in the traffic pattern. The residential component takes temperature. The confounder component, present in the urban region, takes the latent $U_C$, written as $U_C[t] = A[\mathrm{day}(t)]\cdot\,g(t)$, in which $g$ is a fixed within-day shape and $A$ is a daily intensity. The shape $g(t)$ is taken from the low-level gating function. What remains unknown is the intensity, one value per day, estimated jointly with the network weights in Algorithm~\ref{alg:fit}.

\subsection{Fitting \texorpdfstring{$\widehat{\mathcal{M}}_H(\bm{\theta})$}{M\_H(theta)} for $\mathcal{L}_1$-consistency}
The model uses one network per node of $\mathcal{G}_H$ over that node's parent set. The parameters include the sub-network weights as well as the confounder $A[\mathrm{day}]$, which are estimated by minimising the reconstruction error of the observed trajectory $\mathbf{Z}$. Each sub-network is a single-hidden-layer multilayer perceptron (MLP) with ReLU activation and a linear output layer. The optimisation uses the real observed parent values rather than a previous prediction, ensuring that the $\mathcal{L}_1$-consistency is assessed hour by hour against the low-level trajectory, not against a rolled-forward forecast. The weights and the confounder (i.e., $\bm{\theta}$) are optimised in alternating steps. With $\hat{A}[\mathrm{day}]$ held fixed in the confounder slot, the network weights are refitted. With the weights then held fixed, $\hat{A}[\mathrm{day}]$ is searched over 36 points spanning $[-0.5, 2.0]$, taking for each day the single value that minimises the summed reconstruction error across every hour of that day, for the urban NO\textsubscript{x} and PM\textsubscript{2.5} jointly. The two steps are iterated until it fails to improve for eight consecutive rounds, and the fitted model $\widehat{\mathcal{M}}_H(\bm{\theta})$ is secured.

The observations are partitioned into 400 independent week-long realisations of $\mathcal{M}_L$ for training and 25 for testing. Two high-level models are compared. The confounder-unaware model carries no confounder input in its urban NO\textsubscript{x} and PM\textsubscript{2.5}, and its causal diagram $\mathcal{G}_H$ omits the bidirected edge entirely. The confounder-aware model includes the shared latent variable $A[\mathrm{day}]$. Table~\ref{tab:fit-quality} reports the reconstruction loss, which is the variance-normalised mean squared error of the 6 concentrations predicted against the observations. The performance of the confounder-aware model is slightly better: the test loss falls by roughly a factor of three for urban NO\textsubscript{x} and by roughly 1.7 for urban PM\textsubscript{2.5}. Held-out realisations suggest that the confounder carries real predictive signal, obtained before any residual analysis. Test loss exceeds training loss at no node in either model. The recovered $\hat{A}[\mathrm{day}]$ correlates with the true confounder at $r = 0.981$ at the day level over the working hours, although the model never observes it.

\begin{table}[htbp]
\centering
\caption{Reconstruction loss on the training and test partitions}
\label{tab:fit-quality}
\small
\begin{tabular}{@{}llrr@{}}
\toprule
Model & Node & Training & Test \\
\midrule
both & SO\textsubscript{2}, Urban      & 0.01355 & 0.01229 \\
both & NO\textsubscript{x}, Suburban   & 0.00488 & 0.00430 \\
both & SO\textsubscript{2}, Suburban   & 0.00264 & 0.00249 \\
both & PM\textsubscript{2.5}, Suburban & 0.00032 & 0.00029 \\
\addlinespace
\midrule
unaware & NO\textsubscript{x}, Urban   & 0.01358 & 0.01156 \\
aware   & NO\textsubscript{x}, Urban   & \textbf{0.00485} & \textbf{0.00385} \\
unaware & PM\textsubscript{2.5}, Urban & 0.00088 & 0.00077 \\
aware   & PM\textsubscript{2.5}, Urban & \textbf{0.00054} & \textbf{0.00046} \\
\bottomrule
\end{tabular}

\vspace{4pt}
\begin{minipage}{\textwidth}
\footnotesize Note: \emph{unaware} and \emph{aware} abbreviate the confounder-unaware and confounder-aware models.
\end{minipage}
\end{table}

\begin{figure}[htbp]
    \centering
    \includegraphics[width=1\linewidth]{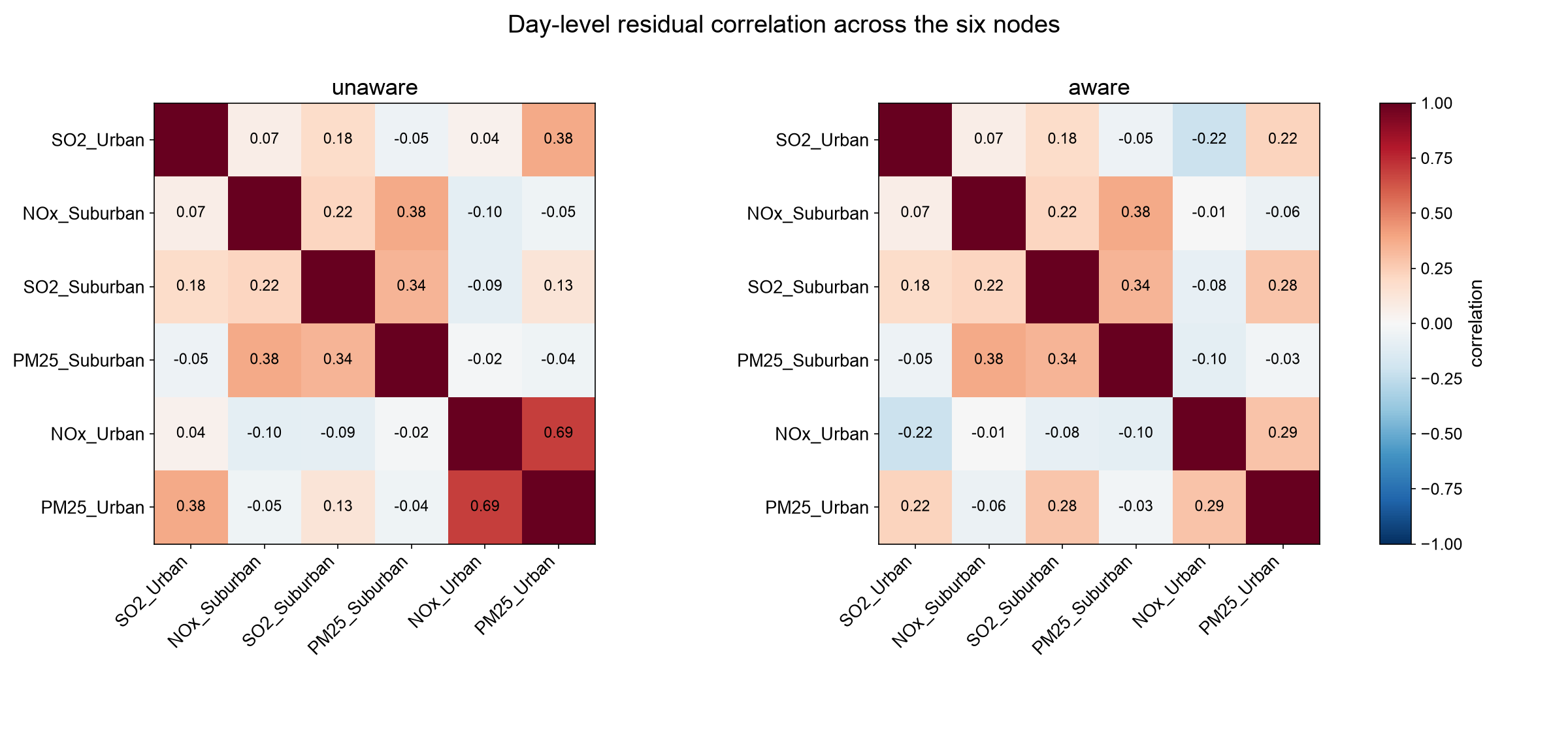}
    \caption{Day-level residual correlation heatmap across the six concentration variables of $\mathcal{G}_H$, under the confounder-unaware model (left) and the confounder-aware model (right), on the held-out test partition.}
    \label{fig:heatmap}
\end{figure}

The residual analysis of Algorithm~\ref{alg:fit} is applied to both models on the test partition. The residual is computed at every hour of every test realisation, using the observed parent values to ensure the residual isolates the leftover noise from the last hour. For the two variables in the confounder-aware model that read the confounder, the input is each test realisation's own best-fit $\hat{A}[\mathrm{day}]$. Because the confounder acts only during working hours, the residuals are aggregated to one value per variable per day over the hours within that window. All fifteen pairs among the six concentration variables are tested. Significance is assessed by 2000 permutations of one variable's day-level residual sequence against the other's, and a pair is reported where the observed correlation exceeds the resulting null spread at $p < 0.05$. Table \ref{tab:residual-corr} gives the correlation and $p$-value for every pair under both models, and Figure \ref{fig:heatmap} renders the same results as two matrices.

\begin{table}[htbp]
\centering
\caption{Day-level residual correlation for all fifteen pairs under both models.}
\label{tab:residual-corr}
\small
\begin{tabular}{@{}llrrrr@{}}
\toprule
$V$ & $V'$ & $r_{\mathrm{unaware}}$ & $p$ & $r_{\mathrm{aware}}$ & $p$ \\
\midrule
NO\textsubscript{x}, Urban   & PM\textsubscript{2.5}, Urban    & \textbf{0.689} & \textbf{$<$0.0005} & 0.286 & $<$0.0005 \\
SO\textsubscript{2}, Urban   & PM\textsubscript{2.5}, Urban    & 0.382 & $<$0.0005 & 0.224 & 0.0015 \\
NO\textsubscript{x}, Suburban & PM\textsubscript{2.5}, Suburban & 0.382 & $<$0.0005 & 0.382 & $<$0.0005 \\
SO\textsubscript{2}, Suburban & PM\textsubscript{2.5}, Suburban & 0.338 & $<$0.0005 & 0.338 & $<$0.0005 \\
SO\textsubscript{2}, Suburban & PM\textsubscript{2.5}, Urban    & 0.129 & 0.090 & 0.280 & $<$0.0005 \\
NO\textsubscript{x}, Suburban & SO\textsubscript{2}, Suburban   & 0.218 & 0.0025 & 0.218 & 0.0035 \\
SO\textsubscript{2}, Urban   & SO\textsubscript{2}, Suburban   & 0.180 & 0.021 & 0.180 & 0.017 \\
SO\textsubscript{2}, Urban   & NO\textsubscript{x}, Urban      & 0.043 & 0.568 & $-$0.216 & 0.0025 \\
SO\textsubscript{2}, Urban   & NO\textsubscript{x}, Suburban   & 0.068 & 0.374 & 0.068 & 0.375 \\
NO\textsubscript{x}, Suburban & NO\textsubscript{x}, Urban     & $-$0.098 & 0.200 & $-$0.006 & 0.939 \\
SO\textsubscript{2}, Suburban & NO\textsubscript{x}, Urban     & $-$0.091 & 0.237 & $-$0.083 & 0.273 \\
NO\textsubscript{x}, Suburban & PM\textsubscript{2.5}, Urban   & $-$0.047 & 0.532 & $-$0.059 & 0.432 \\
PM\textsubscript{2.5}, Suburban & NO\textsubscript{x}, Urban   & $-$0.018 & 0.801 & $-$0.096 & 0.187 \\
PM\textsubscript{2.5}, Suburban & PM\textsubscript{2.5}, Urban & $-$0.044 & 0.561 & $-$0.027 & 0.713 \\
SO\textsubscript{2}, Urban   & PM\textsubscript{2.5}, Suburban & $-$0.053 & 0.480 & $-$0.053 & 0.489 \\
\bottomrule
\end{tabular}
\end{table}

\begin{figure}[htbp]
    \centering
    \includegraphics[width=0.8\linewidth]{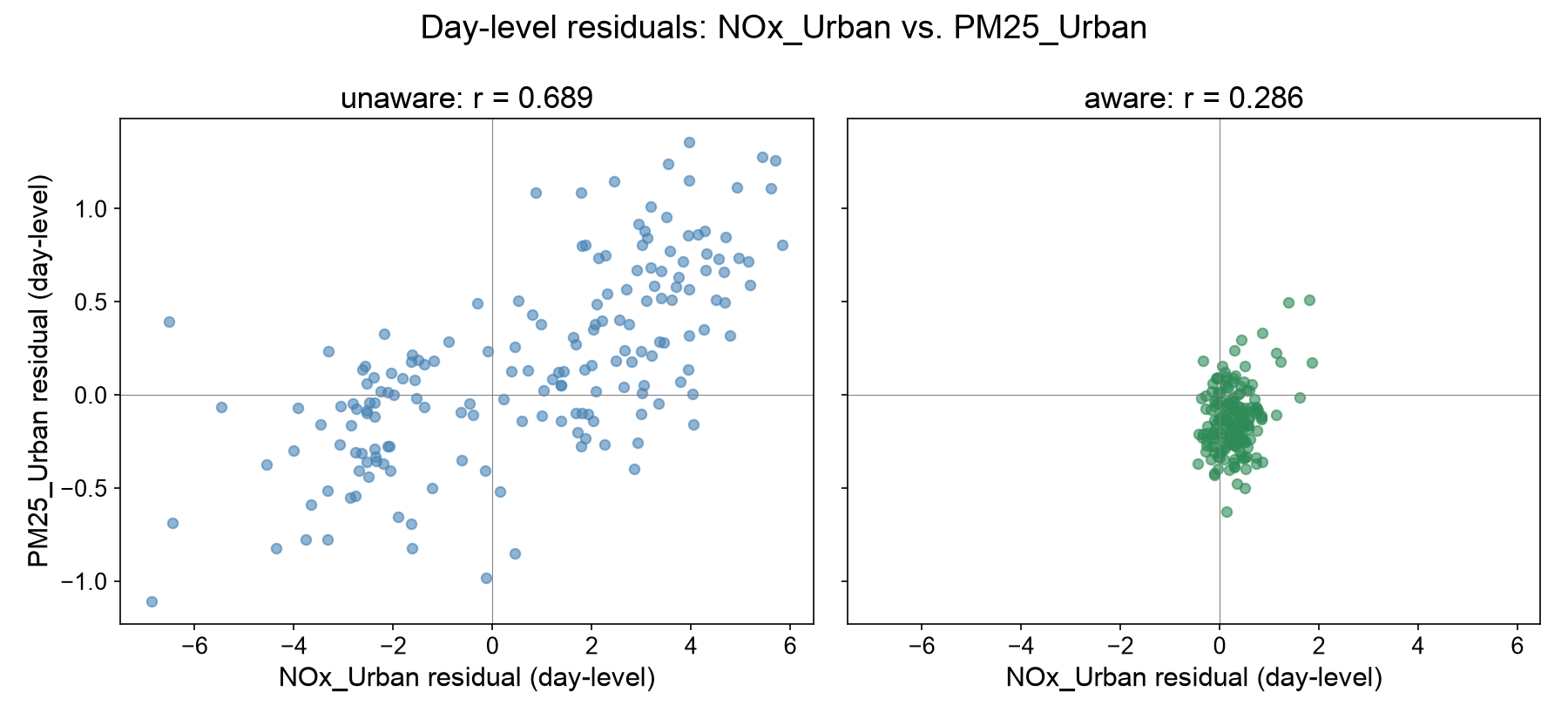}
    \caption{Day-level residuals of urban NO\textsubscript{x} against urban PM\textsubscript{2.5} on the held-out test partition, under the confounder-unaware model (left) and the confounder-aware model (right).}
    \label{fig:residual}
\end{figure}

Under the confounder-unaware model, the pair the confounder acts upon is the most strongly correlated of the fifteen, at $r = 0.689$, and the observed value lies entirely outside the range any of the 2000 reshufflings produced. Under the confounder-aware model, the same correlation falls to $r = 0.286$, with the shared influence now being carried by the confounder rather than left in the residuals. Figure \ref{fig:residual} shows the residual clouds directly: under the unaware model, the day-level residuals lie along a tight diagonal, and under the aware model, the same points loosen.

Six further pairs are also reported in at least one model. The most persistent are the three pairs among the suburban variables, with correlations between 0.22 and 0.38, together with urban SO\textsubscript{2} against urban PM\textsubscript{2.5} and against suburban SO\textsubscript{2}. As Section~\ref{sec:construction} noted, a reported pair calls for reconsideration of the diagram without determining what should change. For instance, the residual of suburban NO\textsubscript{x} still correlates with wind speed, although wind speed is among the inputs to its advection component, which suggests the component may underrepresent the interaction rather than that a confounder is missing.

\subsection{Observational and interventional queries $Q$}
The queries of Section~\ref{sec:queries} are instantiated in the fitted confounder-aware model $\widehat{\mathcal{M}}_H(\bm{\theta}^{\ast})$. Three queries are presented: at $\mathcal{L}_1$, the contribution of each source to urban NO\textsubscript{x} at a given hour; at $\mathcal{L}_2$, a 20\% reduction in traffic emissions, and a ceiling enforcement imposed on urban NO\textsubscript{x}. All three are evaluated on a single illustrative realisation of the test partition, the run with the widest range between its most and least active construction days, comprising five quiet days followed by two of active construction.
 
The observational query asks how much of the urban NO$\text{x}$ concentration at a given hour is attributable to each source: persistence, traffic, residential heating, industrial activity, and the construction confounder. Each is read directly from its corresponding sub-network in $\widehat{\mathcal{M}}_H(\bm\theta^\ast)$, since the model sums every source's own output before producing the total. Specifically, persistence reads only the previous hour's own concentration, traffic reads only the time-of-day features, residential reads only temperature, industrial reads only a constant bias; the confounder reads the day's level of construction activity.

\begin{figure}[!h]
    \centering
    \includegraphics[width=\linewidth]{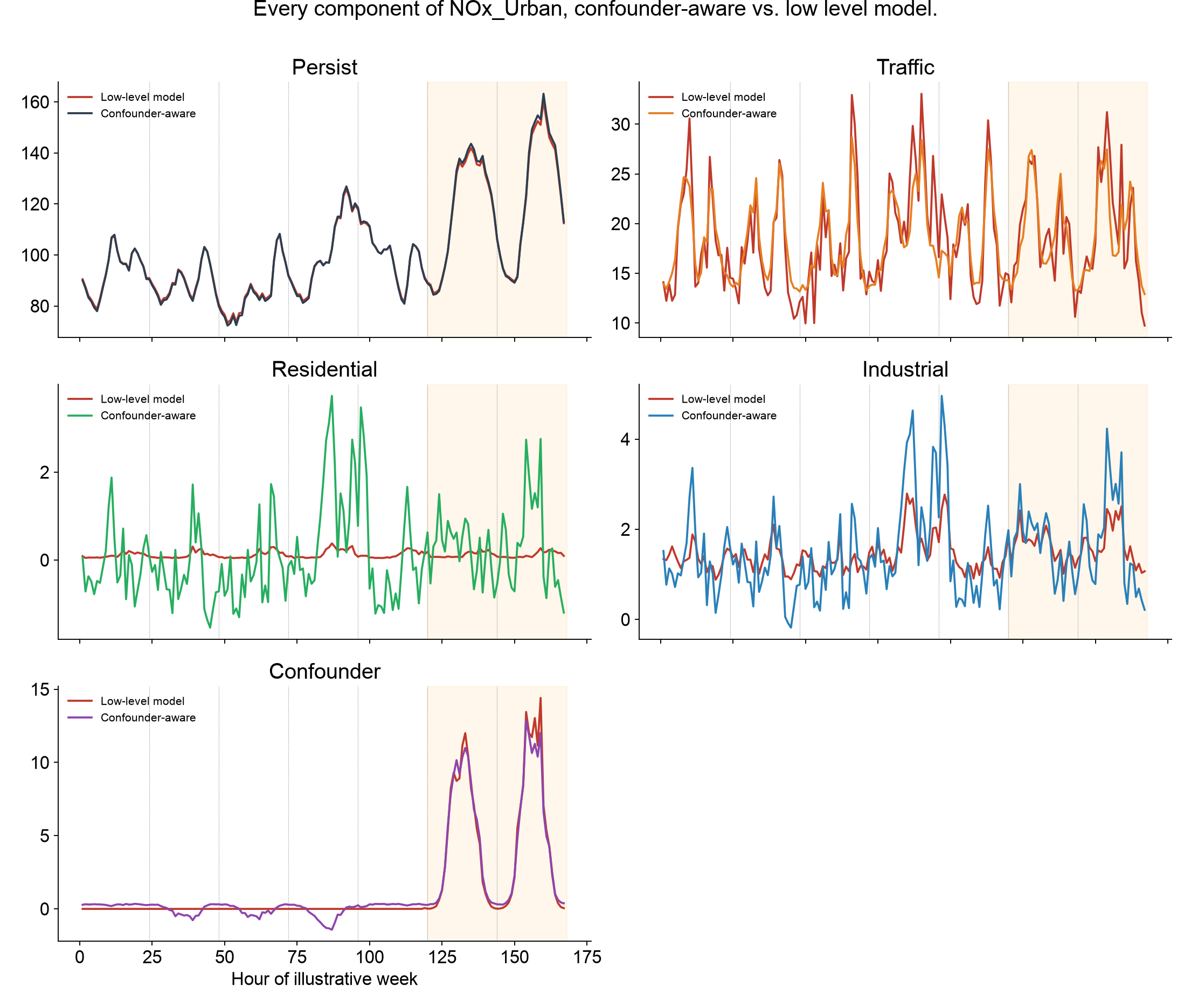}
    \caption{Components estimated for NO\text{x}, Urban across the illustrative test week.}
    \label{fig:nox-components}
\end{figure}

Figure~\ref{fig:nox-components} compares each of the five against its counterpart in $\mathcal{M}_L$ over the illustrative week. Traffic keeps the correct two-peak, morning/evening rush-hour shape (correlation 0.89), but its two-harmonic hour-of-day basis is too coarse to reach the sharper peaks from the low-level model. Residential diverges from the low-level ground truth in shape. Residential's estimate only loosely tracks its temperature-driven pattern, at a correlation of 0.48. This is because the residential contribution to urban NO$_\text{x}$ is small (according to Table \ref{tab:scales}), and the reconstruction loss gives the fit little reason not to park leftover bias on this route. The industrial estimate correlates closely with the true pattern from the low-level model, and its shape comes entirely from the same hour-varying dispersion. The confounder's contribution closely tracks the ground-truth construction pattern, with a correlation of 0.99.

The first interventional query $\mathbb{E}\big[\, C_{t,u}^{s} \;\big|\; \mathrm{do}(\text{traffic scale} = m) \,\big]$, \quad $s \in \{\mathrm{NO_x}, \mathrm{PM_{2.5}}\}$ asks what a 20\% cut to traffic emissions would do to the two urban pollutants. This is the order of change that a congestion charge, a low emission zone, or a partial fleet turnover might deliver, and an assessment of such a measure would normally be requested to estimate its effect in advance. With $m=0.8$, this query is evaluated by scaling the output of the traffic sub-network by $m$ before it enters the additive sum, while holding every other component fixed. 

\begin{figure}[!h]
    \centering
    \includegraphics[width=\linewidth]{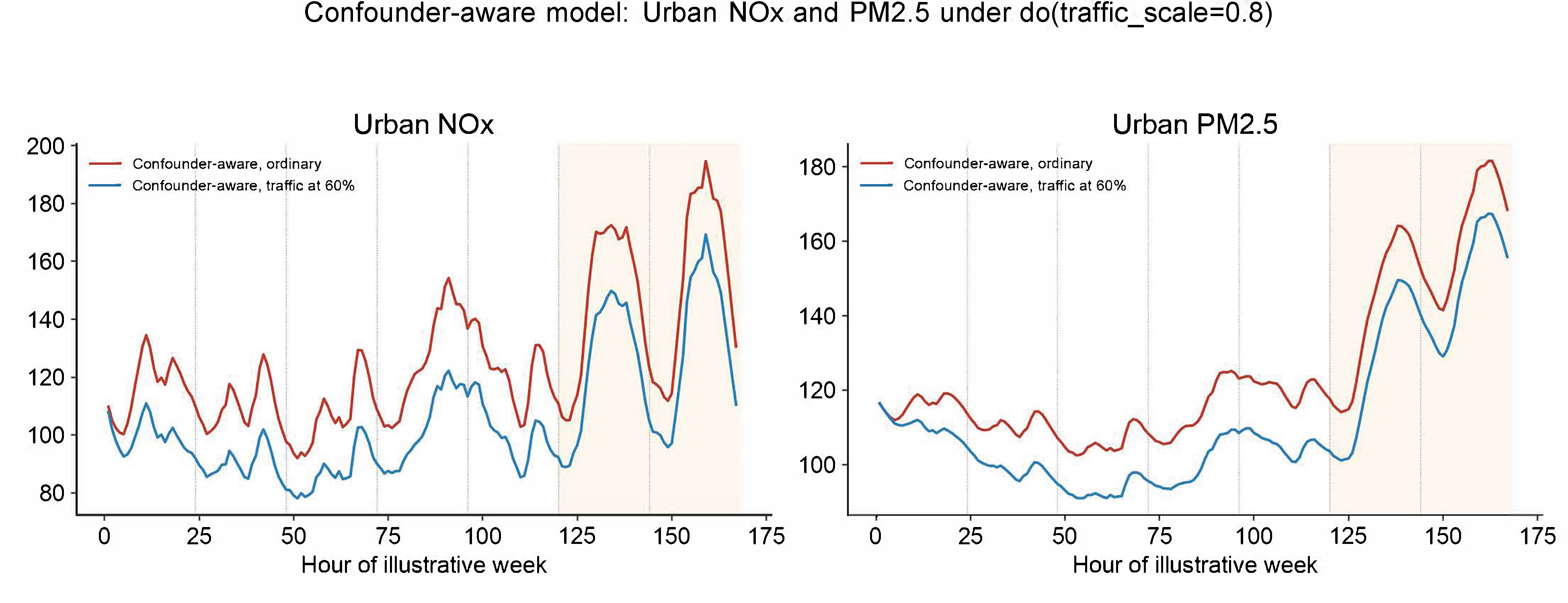}
    \caption{Urban NO\textsubscript{x} and PM\textsubscript{2.5} across the illustrative week, against a 20\% cut to traffic emissions.}
    \label{fig:traffic-scale-effect}
\end{figure}

The direct query result rests on a one-step query, i.e., the model is asked for the effect of the cut on a single hour, with every hour's history held at its observed value. One hour does not show the real effect of the intervention, because cutting emissions in a single hour removes only what would have been emitted that hour, and most of the concentration present comes from earlier hours. The pollutant already in the air has to decay before the reduction is visible. Rolling the model forward under the intervention, Figure~\ref{fig:traffic-scale-effect} gives the trajectory. The separation between the intervened and unintervened runs is negligible at first and widens over the following day as the existing pollutant clears, settling within about 20 to 25 hours. Measured once it settles, the mean reduction is 20.98 ppb (16.6\% of the ordinary mean) for urban NO$\text{x}$ and 12.52 $\mu\text{g/m}^3$ (10.1\% of the ordinary mean) for PM${2.5}$.

Not every query posed is identifiable. The second interventional query, written as $\mathbb{E}\big[\, C_{t,u}^{s} \;\big|\; \mathrm{do}(\text{NO}_\text{x}, \text{Urban}= \min(\text{NO}_\text{x}, \text{Urban}, 138.89)) \,\big]$, \quad $s \in \{\mathrm{PM_{2.5}}\}$, caps urban NO\textsubscript{x} at the 75th percentile of its own distribution over the illustrative week. Specifically, the urban NO\textsubscript{x} is set to its observed value at every hour below that ceiling and pinned to the ceiling at every hour it would exceed it. Physically, this represents an emergency action level used in air-quality regulations: a cap that binds only during the worst pollution episodes and leaves ordinary hours untouched. Even so, the intervention still idealises the mechanism: no policy forces ambient concentration directly, only the emissions that produce it.

The query is structurally non-identifiable. Urban NO\textsubscript{x} and PM\textsubscript{2.5} are the pair of variables joined both by a direct causal edge, i.e., NO\textsubscript{x} enters PM\textsubscript{2.5} via secondary chemistry, and by a bidirected path through the shared confounder $A[\mathrm{day}]$. A variable that is simultaneously a direct cause of, and confounded with, its own outcome is Pearl's classical bow graph, the textbook example of a query that Shpitser and Pearl's criterion rules as non-identifiable from observational data alone \citep{shpitser2008complete}. Given the construction confounder with a bidirected edge in the model, this query is guaranteed to fail in theory.

Algorithm~\ref{alg:id} reaches the same conclusion from the data. The algorithm searches over every parameter setting whose reconstruction loss against the observed data stays within a fixed tolerance, and reports the minimum and maximum the query takes across that admissible set. Where the two disagree, the query cannot be identified from the data. The search returns a range of $[132.93, 153.31]$, a width of 20.38 $\mu\text{g/m}^3$, with both endpoints holding the reconstruction constraint exactly rather than merely within tolerance. The width therefore reflects a feature of the admissible region, and the query is substantially non-identified rather than merely difficult to pin down.

\section{Discussion}
The causal diagram of the synthetic experiment was assembled by hand. Four zones, three species and one reaction admit that treatment; a city does not. A domain of hundreds of links across a dozen sectors, with ten to twenty chemical pathways among the species a sensor network reports, must have its diagram generated rather than drawn, and a knowledge graph is the natural object from which to generate it.

Two kinds of knowledge inform the construction of the causal diagram $\mathcal{G}_H$ and the high-level model $\mathcal{M}_H$, and they enter in different ways. On the one hand, chemical knowledge is schema-level: a reaction relates named species under stated conditions, and the relation holds wherever those species are present. The pathways connect species a sensor network reports, each lumping several elementary reactions of a mechanism. For example, nitrogen dioxide bears on particulate nitrate through oxidation to nitric acid and partitioning with ammonia, and sulfur dioxide bears on particulate sulfate. Because intermediates in these chains are rarely measured, each pathway is carried as an edge rather than a chain. Conversion along such a chain depends on its intermediates, and those intermediates depend in turn on prevailing conditions. For instance, radiation, ozone and water vapour set the hydroxyl radical, and humidity and composition set aerosol liquid water. Therefore, the rate of conversion is governed by radiation, temperature and humidity, although none appears in any individual reaction along the chain. In short, the chemical schema supplies the edges in $\mathcal{G}_H$, asserting which species convert to which. The component that takes those species as arguments supplies the rate as a function of the conditions the intermediates respond to. 

Urban knowledge, on the other hand, is instance-level: it records which sources lie in which cell, which sensors report which species at which location, and how the spatial units nest. Ontologies developed for the built environment, sensor networks, and city models supply this, and they fix the variable inventory of $\mathcal{G}_H$ and the spatial relations among the variables. 

The high-level urban air quality model built this way has three clear advantages. First, the variable inventory becomes machine-readable, which specifies the abstraction map $\tau$. A user can then establish which fine-grained quantity each macro-variable aggregates, instead of inferring it from the surrounding text. Second, the candidate edge set becomes auditable, because every edge traces to an asserted relation and can be shown to whoever maintains the mechanism. Last but not least, the construction of $\mathcal{M}_H$ becomes portable. The chemical schema is common to any city, and a diagram generated from it transfers to a second domain by substituting only the instance-level graph. However, a knowledge graph records what a city knows about itself, and no city knows everything that affects its air. An ontology can include the road network, permitted industrial sites, and building stock, yet it will still omit activities nobody registers, e.g., an unrecorded construction site, a diverted freight route, or a period of unusual domestic burning. The confounder is how such omitted influences enter the model: an unmeasured quantity affecting two or more variables is represented by a bidirected edge. Algorithm~\ref{alg:fit} indicates where the diagram requires a bidirected edge, in order to ensure the model represents the city more faithfully than its records do.

Artificial intelligence plays an increasingly important role in urban air quality analysis. Deep networks forecast pollutant fields with high fidelity: graph neural networks that treat monitoring sites as connected nodes now give competitive short-range forecasts of particulate matter \citep{wang2020pm2}, and neural emulators reproduce the behaviour of a full chemical mechanism at a fraction of the cost \citep{kelp2022online}. More significantly, a single foundation model fine-tuned on atmospheric-composition data can produce five-day global air-pollution forecasts that match or outperform an operational chemistry model on almost three-quarters of targets, at a tiny fraction of its computational cost \citep{bodnar2025foundation}. If a model's purpose were to predict the next state of the air, existing technologies would be more than enough.

A deep network converts its inputs into an embedding or latent representation: a point in a high-dimensional space, from which further arithmetic reads off the answer. Recent work finds that high-level ideas are often carried as directions within the embedding space \citep{park2023linear}, and that many such directions can be extracted and named \citep{templeton2026scaling}. But a direction in an embedding is not the same thing as a variable with a mechanism. Nothing fixes which fine-grained quantity it represents, over which streets, or over which averaging time. Because a network can pack more features than it has dimensions, i.e., the factors of variation overlap, one cannot alter the emissions of one source while holding dispersion and chemistry fixed. This matters in air quality precisely because the domain punishes the confusion of prediction with intervention. When the 2020 lockdowns cut traffic sharply, a model tuned to ordinary conditions would have predicted nitrogen dioxide and ozone to fall together. Instead, deweathered observations across eleven cities showed nitrogen dioxide falling by less than the emission cut implied and ozone rising, with total oxidant barely changed \citep{shi2021abrupt}. The sign of the ozone response is not an accident of that year: near traffic, ozone sits in a regime where removing nitrogen oxides releases it from titration, and the emission-concentration relationship depends on the regime. In this case, the failure is due to the absence of a variable one can intervene upon.

This paper offers a way to make reasoning over such coarse variables actually happen. Admissibility checks the variables against physical scales, ensuring a cell-average concentration is treated as a variable only over an interval long enough for sub-cell processes to settle yet short enough that transport has not left the cell. The idea of valid coarse descriptions originates from the lumpability of aggregated Markov processes to causal emergence \citep{hoel2013quantifying}, but here the condition is read from the physics rather than discovered from data. Learning the causal diagram from chemistry and geometry ensures the edges have a warrant that does not depend on the data fitting. The identifiability check allows the model to decline: faced with a query that the structure and the evidence do not determine, it returns ``not identifiable" rather than a confident number.

This opens a new thread on how AI reasons over abstracted concepts in air quality. The conversational AI agents now being built to answer questions about urban air reason over embeddings \citep{xie2026synthesising}. A high-level model in this paper offers such an AI system a different footing: a physics-grounded account of what its variables represent, where its structure came from, and which questions that structure entitles it to answer. What it supplies is warrant: an answer that holds for the city, the sources and the observations actually in question rather than for a generic case, together with a statement of what data can determine and what they cannot.

\section{Conclusion}
Urban air quality is typically analysed over abstracted variables (e.g., zones, fleets, and hourly means), and such abstraction should reflect the physical mechanisms with fidelity under the interventions it is used to assess, whether emission controls, traffic measures, or fleet change. This paper examines what such an abstraction must satisfy before it may be used for abduction and intervention, and locates the answer in three places. The first is the choice of spatial and temporal abstraction: a cell average constitutes a causal variable only over an interval long enough for the processes it collapses to have settled and short enough that influence has not left the cell. In the cases where no interval satisfies both, the variables must be redefined in a source–receptor formulation. The second is the causal structure: the causal diagram, supplied by the chemical mechanism and the geometry of the domain, fixes which relations exist among the variables; the observations determine the strength and sign of each relation, and residual independence after fitting tests whether a confounder is omitted. The third is identifiability, which is decided query by query: whether the diagram and the observations settle an interventional question is established before any answer is returned.

The controlled synthetic experiment exercises the proposed approach. Construction activity that a model never observed left its signature in the residuals: under the confounder-unaware model, the day-level residuals of urban NO\textsubscript{x} and PM\textsubscript{2.5} correlated at 0.689, and the correlation fell to 0.286 once the confounder was carried as a latent daily intensity. With the confounder in place, the effects of a 20\% cut in traffic emissions was identifiable, lowering urban NO\textsubscript{x} by 20.98 ppb and PM\textsubscript{2.5} by 12.52 $\mu\text{g/m}^3$. The effects of imposing a ceiling on urban NO\textsubscript{x} at its 75th percentile were not identifiable: between models that fit the observations equally well, the confounding left a gap of 20.38~$\mu\text{g/m}^3$ between the maximum and minimum PM\textsubscript{2.5} the query could take, and the query was declined accordingly.

For AI reasoning about urban air, this construction marks a path toward systems grounded in physics rather than fit; specific to the city and scenario in question rather than a generic case; and aware of their own limits, declining questions the evidence cannot settle. 

\bibliographystyle{unsrtnat}  
\bibliography{references}

\end{document}

%% file: Figures/Causal_abstraction.tex
\begin{figure}[htbp]
\centering
\begin{tikzpicture}[
  every node/.style={font=\small},
  ar/.style={-{Stealth[length=1.8mm]}},
  dar/.style={-{Stealth[length=1.8mm]}, dashed}
]

\node (L2L) at (0,1.8) {$\mathcal{L}_2(\mathcal{M}_L)$};
\node (L2H) at (8,1.8) {$\mathcal{L}_2(\mathcal{M}_H)$};
\node (L1L) at (0,0)   {$\mathcal{L}_1(\mathcal{M}_L)$};
\node (L1H) at (8,0)   {$\mathcal{L}_1(\mathcal{M}_H)$};

\draw[dar] (L2L) -- node[above] {$\tau$}
                    node[below, font=\footnotesize] {Interventional consistency} (L2H);
\draw[ar]  (L1L) -- node[above] {$\tau$}
                    node[below, font=\footnotesize] {Admissibility and $\mathcal{L}_1$-consistency} (L1H);

\draw[ar] (L1L) -- node[left] {$\mathrm{do}(i)$} (L2L);
\draw[ar] (L1H) -- node[right] {$\mathrm{do}\big(\omega(i)\big)$} (L2H);

\end{tikzpicture}

\caption{The abstraction relation. The lower edge holds where $\tau$ is admissible and $\mathcal{M}_H$ is fitted to the observations. The upper edge is what an interventional query $Q$ requires, and whether it holds for a given $Q$ is a question of identifiability.}
\label{fig:causal}
\end{figure}

%% file: Figures/Conceptual_system.tex
\begin{figure}[htbp]
\centering
\begin{tikzpicture}[
  zone/.style = {draw, thick, rounded corners=3pt,
                 minimum width=1.9cm, minimum height=1.0cm,
                 align=center, font=\footnotesize},
  emis/.style = {draw, line width=0.5pt, rounded corners=2pt,
                 minimum width=1.3cm, minimum height=0.5cm,
                 align=center, font=\scriptsize},
  hid/.style  = {draw, line width=0.7pt, dashed, rounded corners=3pt,
                 minimum width=2.0cm, minimum height=0.6cm,
                 align=center, font=\scriptsize, orange!85!black,
                 text=orange!85!black},
  flow/.style = {-{Stealth[length=2mm]}, line width=0.7pt},
  wind/.style = {-{Stealth[length=3.5mm, width=3mm]}, line width=1.6pt,
                 gray!45},
  conf/.style = {-{Stealth[length=1.8mm]}, line width=0.6pt, dashed,
                 orange!85!black},
  lab/.style  = {font=\scriptsize},
  clab/.style = {font=\scriptsize, orange!85!black}
]
 
\node[zone] (ind) at (0,0)   {Industrial};
\node[zone] (tra) at (3.8,0) {Traffic};
\node[zone] (res) at (7.6,0) {Residential};
\node[zone] (sub) at (11.4,0) {Suburban};
 
\draw[flow] (ind) -- node[lab, above, inner sep=2pt] {\textit{advection}} (tra);
\draw[flow] (tra) -- node[lab, above, inner sep=2pt] {\textit{advection}} (res);
\draw[flow] (res) -- node[lab, above, inner sep=2pt] {\textit{advection}} (sub);
 
\node[emis] (eind) at (0,-1.5)   {\textit{emissions}};
\node[emis] (etra) at (3.8,-1.5) {\textit{emissions}};
\node[emis] (eres) at (7.6,-1.5) {\textit{emissions}};
\node[emis] (esub) at (11.4,-1.5) {\textit{emissions}};
 
\foreach \e/\z in {eind/ind, etra/tra, eres/res, esub/sub}{
  \draw[flow] (\e) -- (\z);
}
 
\draw[wind] (-0.4,1.5) -- (11.3,1.5);
\node[lab, gray!60!black] at (5.45,1.88) {\textit{Prevailing wind}};
 
\node[hid] (u) at (1.85,-3.1) {construction activity};
 
\draw[conf] (u.north west) to[out=110, in=-90] (eind.south);
\draw[conf] (u.north east) to[out=70,  in=-90] (etra.south);
\node[clab, anchor=north] at (1.85,-2.25) {via emissions};
 
\draw[conf] (u.west) to[out=180, in=-120] (ind.south west);
\draw[conf] (u.east) to[out=0,   in=-60]  (tra.south east);
\node[clab, anchor=north west, align=left] at (4.65,-2.05)
  {direct dust,\\bypasses emissions};
 
\end{tikzpicture}

\caption{Layout of the conceptual system. Four zones lie along a fixed prevailing wind, each affected by its own emissions and by inflow from the zone immediately upwind. Construction activity reaches the industrial and traffic zones in two ways: through their emissions, and as dust entering PM\textsubscript{2.5} directly.}
\label{fig:layout}
\end{figure}

%% file: Figures/Causal_diagram.tex

\begin{figure}[ht]
\centering
\begin{tikzpicture}[
  nd/.style   = {draw, circle, minimum size=0.78cm, inner sep=0pt,
                 font=\tiny, align=center, fill=white},
  gh/.style   = {draw, circle, minimum size=0.78cm, inner sep=0pt,
                 font=\tiny, align=center, fill=white,
                 gray!50, text=gray!55!black},
  drv/.style  = {draw, circle, minimum size=0.30cm, inner sep=0pt,
                 font=\tiny, align=center, fill=white, line width=0.3pt},
  lat/.style  = {draw, dashed, circle, minimum size=0.8cm, inner sep=0pt,
                 font=\tiny, align=center, orange!85!black,
                 text=orange!85!black, fill=white},
  reg/.style  = {draw, dashed, line width=0.5pt, rounded corners=5pt,
                 inner sep=10pt},
  regf/.style = {draw, dashed, line width=0.4pt, rounded corners=8pt,
                 inner sep=8pt, gray!45},
  ed/.style   = {-{Stealth[length=1.5mm]}, line width=0.55pt},
  adv/.style  = {-{Stealth[length=1.4mm]}, line width=0.5pt, blue!55!black},
  led/.style  = {-{Stealth[length=1.5mm]}, line width=0.55pt, dashed,
                 orange!85!black},
  fade/.style = {-{Stealth[length=1.3mm]}, line width=0.4pt, gray!55},
  dvd/.style  = {-{Stealth[length=1.1mm]}, line width=0.35pt},
  clab/.style = {font=\scriptsize},
  rlab/.style = {font=\scriptsize}
]

\def\cP{-5.4} \def\cT{0} \def\cN{5.4}
\def\yU{1.7} \def\yS{-1.5}
\def\off{1.05}

\node[nd] (A)  at (\cT,4.9) {$A$};
\node[nd] (E)  at (\cT,3.4) {$E$};
\node[nd] (US) at (\cT-1.3,\yU) {SO$_2$};
\node[nd] (UN) at (\cT,\yU)     {NO$_x$};
\node[nd] (UP) at (\cT+1.3,\yU) {PM};
\node[nd] (SS) at (\cT-1.3,\yS) {SO$_2$};
\node[nd] (SN) at (\cT,\yS)     {NO$_x$};
\node[nd] (SP) at (\cT+1.3,\yS) {PM};
\node[lat] (U) at (\cT-3.1,4.9) {$U_C$};
\node[font=\scriptsize\bfseries, orange!85!black, anchor=east] at ($(U.west)+(-0.15,0)$) {Confounder};

\foreach \n in {US, UN, UP, SS, SN, SP, E}{
  \node[drv] (w\n) at ($(\n)+(0.50,0.50)$) {$W$};
}

\node[gh] (USp) at (\cP-1.3,\yU+\off) {SO$_2$};
\node[gh] (UNp) at (\cP,\yU+\off)     {NO$_x$};
\node[gh] (UPp) at (\cP+1.3,\yU+\off) {PM};
\node[gh] (SSp) at (\cP-1.3,\yS+\off) {SO$_2$};
\node[gh] (SNp) at (\cP,\yS+\off)     {NO$_x$};
\node[gh] (SPp) at (\cP+1.3,\yS+\off) {PM};

\node[gh] (USn) at (\cN-1.3,\yU-\off) {SO$_2$};
\node[gh] (UNn) at (\cN,\yU-\off)     {NO$_x$};
\node[gh] (UPn) at (\cN+1.3,\yU-\off) {PM};
\node[gh] (SSn) at (\cN-1.3,\yS-\off) {SO$_2$};
\node[gh] (SNn) at (\cN,\yS-\off)     {NO$_x$};
\node[gh] (SPn) at (\cN+1.3,\yS-\off) {PM};

\begin{scope}[on background layer]
  \node[reg,  fit=(US)(UP)(wUS)(wUP)] (urb) {};
  \node[reg,  fit=(SS)(SP)(wSS)(wSP)] (sub) {};
  \node[regf, fit=(USp)(UPp)] {};
  \node[regf, fit=(SSp)(SPp)] {};
  \node[regf, fit=(USn)(UPn)] {};
  \node[regf, fit=(SSn)(SPn)] {};
\end{scope}
\node[rlab, anchor=east] at ($(urb.west)+(-0.12,0)$) {Urban};
\node[rlab, anchor=east] at ($(sub.west)+(-0.12,0)$) {Suburban};

\node[clab] at (\cT,5.8) {$t$};
\node[clab, gray!55!black] at (\cP,\yU+\off+1.3) {$t-1$};
\node[clab, gray!55!black] at (\cN,\yU-\off+1.3) {$t+1$};

\begin{scope}[on background layer]

  \foreach \n in {US, UN, UP, SS, SN, SP, E}{
    \draw[dvd] (w\n) -- (\n);
  }

  \draw[ed] (A) -- (E);
  \foreach \n in {US, UN, UP, SS, SN, SP}{ \draw[ed] (E) -- (\n); }
  \draw[ed] (UN) -- (UP);
  \draw[ed] (US) to[out=-40,in=-140] (UP);
  \draw[ed] (SN) -- (SP);
  \draw[ed] (SS) to[out=-40,in=-140] (SP);

  \draw[led] (U) to[out=25,in=145] (E);
  \draw[led] (U) to[out=-80,in=170] (UP);

  \foreach \p/\q in {USp/US, UNp/UN, UPp/UP, SSp/SS, SNp/SN, SPp/SP,
                     US/USn, UN/UNn, UP/UPn, SS/SSn, SN/SNn, SP/SPn}{
    \draw[fade] (\p) -- (\q);
  }
  \foreach \p/\q in {USp/SS, UNp/SN, UPp/SP, US/SSn, UN/SNn, UP/SPn}{
    \draw[adv] (\p) to[out=-30,in=160] (\q);
  }

\end{scope}

\end{tikzpicture}
\caption{The high-level diagram $\mathcal{G}_H$ at hour $t$, with $t-1$ and $t+1$ in grey. Activity $A$ is an endogenous root, and emissions $E$ are not observed. Meteorology is shown as $W$ beside each node that takes it, including the dispersion index, wind speed and temperature. Black edges are the mechanisms at $t$, grey edges are persistence, and blue edges are advection from the urban to the suburban region at lag one. The construction activity $U_C$ is exogenous, entering emissions and PM\textsubscript{2.5} through separate paths.}
\label{fig:GH}
\end{figure}

%% file: references.bib
@article{rubenstein2017causal,
  title={Causal consistency of structural equation models},
  author={Rubenstein, Paul K and Weichwald, Sebastian and Bongers, Stephan and Mooij, Joris M and Janzing, Dominik and Grosse-Wentrup, Moritz and Sch{\"o}lkopf, Bernhard},
  journal={arXiv preprint arXiv:1707.00819},
  year={2017}
}

@inproceedings{beckers2019abstracting,
  title={Abstracting causal models},
  author={Beckers, Sander and Halpern, Joseph Y},
  booktitle={Proceedings of the AAAI Conference on Artificial Intelligence},
  volume={33},
  number={01},
  pages={2678--2685},
  year={2019}
}

@article{scholkopf2021toward,
  title={Toward causal representation learning},
  author={Sch{\"o}lkopf, Bernhard and Locatello, Francesco and Bauer, Stefan and Ke, Nan Rosemary and Kalchbrenner, Nal and Goyal, Anirudh and Bengio, Yoshua},
  journal={Proceedings of the IEEE},
  volume={109},
  number={5},
  pages={612--634},
  year={2021},
  publisher={IEEE}
}

@article{geiger2021causal,
  title={Causal abstractions of neural networks},
  author={Geiger, Atticus and Lu, Hanson and Icard, Thomas and Potts, Christopher},
  journal={Advances in Neural Information Processing Systems},
  volume={34},
  pages={9574--9586},
  year={2021}
}

@article{xia2021causal,
  title={The causal-neural connection: Expressiveness, learnability, and inference},
  author={Xia, Kevin and Lee, Kai-Zhan and Bengio, Yoshua and Bareinboim, Elias},
  journal={Advances in Neural Information Processing Systems},
  volume={34},
  pages={10823--10836},
  year={2021}
}

@article{chalupka2015visual,
  title={Visual causal feature learning},
  author={Chalupka, Krzysztof and Perona, Pietro and Eberhardt, Frederick},
  journal={Proceedings of the Thirty-First Conference on Uncertainty in Artificial Intelligence},
  pages={181--190},
  year={2015}
}

@article{garcez2023neurosymbolic,
  title={Neurosymbolic {AI}: the 3rd wave},
  author={Garcez, Artur d’Avila and Lamb, Luis C},
  journal={Artificial Intelligence Review},
  volume={56},
  number={11},
  pages={12387--12406},
  year={2023},
  publisher={Springer}
}

@article{sin2023cfd,
  title={{CFD} modeling on the canyon ventilation and pollutant exposure in asymmetric street canyons with continuity/discontinuity balconies},
  author={Sin, Chung Hyok and Cui, Peng-Yi and Luo, Yang and Jon, Kwang Song and Huang, Yuan-dong},
  journal={Atmospheric Pollution Research},
  volume={14},
  number={1},
  pages={101641},
  year={2023},
  publisher={Elsevier}
}

@article{xie2024improvement,
  title={Improvement of the ozone forecast over Beijing through combining the chemical transport model with multiple machine learning methods},
  author={Xie, Jiali and Tang, Xiao and Zheng, Fei and Wang, Xidong and Ding, Ning and Song, Yating and Luo, Hongyan and Wu, Huangjian and Kong, Lei and Zhu, Jiang and others},
  journal={Atmospheric Pollution Research},
  volume={15},
  number={8},
  pages={102184},
  year={2024},
  publisher={Elsevier}
}

@article{han2022semi,
  title={Semi-supervised air quality forecasting via self-supervised hierarchical graph neural network},
  author={Han, Jindong and Liu, Hao and Xiong, Haoyi and Yang, Jing},
  journal={IEEE Transactions on Knowledge and Data Engineering},
  volume={35},
  number={5},
  pages={5230--5243},
  year={2022},
  publisher={IEEE}
}

@article{beelen2013development,
  title={Development of {NO2} and {NOx} land use regression models for estimating air pollution exposure in 36 study areas in Europe--The {ESCAPE} project},
  author={Beelen, Rob and Hoek, Gerard and Vienneau, Danielle and Eeftens, Marloes and Dimakopoulou, Konstantina and Pedeli, Xanthi and Tsai, Ming-Yi and K{\"u}nzli, Nino and Schikowski, Tamara and Marcon, Alessandro and others},
  journal={Atmospheric Environment},
  volume={72},
  pages={10--23},
  year={2013},
  publisher={Elsevier}
}

@inproceedings{wang2020pm2,
  title={{PM2.5-GNN}: A domain knowledge enhanced graph neural network for {PM2.5} forecasting},
  author={Wang, Shuo and Li, Yanran and Zhang, Jiang and Meng, Qingye and Meng, Lingwei and Gao, Fei},
  booktitle={Proceedings of the 28th International Conference on Advances in Geographic Information Systems},
  pages={163--166},
  year={2020}
}

@article{zevcevic2021relating,
  title={Relating graph neural networks to structural causal models},
  author={Ze{\v{c}}evi{\'c}, Matej and Dhami, Devendra Singh and Veli{\v{c}}kovi{\'c}, Petar and Kersting, Kristian},
  journal={arXiv preprint arXiv:2109.04173},
  year={2021}
}

@article{luecken2019multipollutant,
  title={Multipollutant modeling of ozone, reactive nitrogen and {HAPs} across the continental {US} with {CMAQ-CB6}},
  author={Luecken, DJ and Yarwood, G and Hutzell, WT},
  journal={Atmospheric Environment},
  volume={201},
  pages={62--72},
  year={2019},
  publisher={Elsevier}
}

@article{nenes2020aerosol,
  title={Aerosol {pH} and liquid water content determine when particulate matter is sensitive to ammonia and nitrate availability},
  author={Nenes, Athanasios and Pandis, Spyros N and Weber, Rodney J and Russell, Armistead},
  journal={Atmospheric Chemistry and Physics},
  volume={20},
  number={5},
  pages={3249--3258},
  year={2020},
  publisher={Copernicus Publications G{\"o}ttingen, Germany}
}

@article{guo2018effectiveness,
  title={Effectiveness of ammonia reduction on control of fine particle nitrate},
  author={Guo, Hongyu and Otjes, Rene and Schlag, Patrick and Kiendler-Scharr, Astrid and Nenes, Athanasios and Weber, Rodney J},
  journal={Atmospheric Chemistry and Physics},
  volume={18},
  number={16},
  pages={12241--12256},
  year={2018},
  publisher={Copernicus Publications G{\"o}ttingen, Germany}
}

@article{spirtes2004causal,
  title={Causal inference of ambiguous manipulations},
  author={Spirtes, Peter and Scheines, Richard},
  journal={Philosophy of Science},
  volume={71},
  number={5},
  pages={833--845},
  year={2004},
  publisher={Cambridge University Press}
}

@article{grange2019using,
  title={Using meteorological normalisation to detect interventions in air quality time series},
  author={Grange, Stuart K and Carslaw, David C},
  journal={Science of the Total Environment},
  volume={653},
  pages={578--588},
  year={2019},
  publisher={Elsevier}
}

@article{zhang2025adjoint,
  title={Adjoint analysis of {PM2.5} and {O3} episodes in priority control zones in China},
  author={Zhang, Ruixin and Chen, Zhihong and Wu, Xueyan and Liu, Qiming and Mai, Zelin and Zheng, Zhiyu and Chen, Yilin and Tao, Shu and Hu, Yongtao and Zhao, Shunliu and others},
  journal={Environmental Science and Ecotechnology},
  pages={100612},
  year={2025},
  publisher={Elsevier}
}

@article{khamlich2023physics,
  title={A physics-based reduced order model for urban air pollution prediction},
  author={Khamlich, Moaad and Stabile, Giovanni and Rozza, Gianluigi and K{\"o}rnyei, L{\'a}szl{\'o} and Horv{\'a}th, Zolt{\'a}n},
  journal={Computer Methods in Applied Mechanics and Engineering},
  volume={417},
  pages={116416},
  year={2023},
  publisher={Elsevier}
}

@article{raissi2019physics,
  title={Physics-informed neural networks: A deep learning framework for solving forward and inverse problems involving nonlinear partial differential equations},
  author={Raissi, Maziar and Perdikaris, Paris and Karniadakis, George E},
  journal={Journal of Computational physics},
  volume={378},
  pages={686--707},
  year={2019},
  publisher={Elsevier}
}

@article{shi2025phy,
  title={{Phy-APMR}: A physics-informed air pollution map reconstruction approach with mobile crowd-sensing for fine-grained measurement},
  author={Shi, Rongye and Luo, Ji and Zhou, Nan and Liu, Yuxuan and Hong, Chaopeng and Zhang, Xiao-Ping and Chen, Xinlei},
  journal={Building and Environment},
  volume={272},
  pages={112634},
  year={2025},
  publisher={Elsevier}
}

@article{amann2011cost,
  title={Cost-effective control of air quality and greenhouse gases in Europe: Modeling and policy applications},
  author={Amann, Markus and Bertok, Imrich and Borken-Kleefeld, Jens and Cofala, Janusz and Heyes, Chris and H{\"o}glund-Isaksson, Lena and Klimont, Zbigniew and Nguyen, Binh and Posch, Maximilian and Rafaj, Peter and others},
  journal={Environmental Modelling \& Software},
  volume={26},
  number={12},
  pages={1489--1501},
  year={2011},
  publisher={Elsevier}
}

@article{hopke2016review,
  title={Review of receptor modeling methods for source apportionment},
  author={Hopke, Philip K},
  journal={Journal of the Air \& Waste Management Association},
  volume={66},
  number={3},
  pages={237--259},
  year={2016},
  publisher={Taylor \& Francis}
}

@article{clappier2017source,
  title={Source apportionment and sensitivity analysis: two methodologies with two different purposes},
  author={Clappier, Alain and Belis, Claudio A and Pernigotti, Denise and Thunis, Philippe},
  journal={Geoscientific Model Development},
  volume={10},
  number={11},
  pages={4245--4256},
  year={2017},
  publisher={Copernicus Publications G{\"o}ttingen, Germany}
}

@article{xia2023neural,
  title={Neural causal models for counterfactual identification and estimation},
  author={Xia, Kevin and Pan, Yushu and Bareinboim, Elias},
  journal={Proceedings of the 11th International Conference on Learning Representations},
  year={2023}
}

@incollection{bareinboim2022pearl,
  title={On {Pearl’s} hierarchy and the foundations of causal inference},
  author={Bareinboim, Elias and Correa, Juan D and Ibeling, Duligur and Icard, Thomas},
  booktitle={Probabilistic and causal inference: the works of judea pearl},
  pages={507--556},
  year={2022}
}

@inproceedings{beckers2020approximate,
  title={Approximate causal abstractions},
  author={Beckers, Sander and Eberhardt, Frederick and Halpern, Joseph Y},
  booktitle={Uncertainty in artificial intelligence},
  booktitle={Proceedings of the 35th Conference on Uncertainty in Artificial Intelligence},
  pages={606--615},
  year={2020},
}

@article{solazzo2007transfer,
  title={Transfer processes in a simulated urban street canyon},
  author={Solazzo, E and Britter, RE},
  journal={Boundary-Layer Meteorology},
  volume={124},
  number={1},
  pages={43--60},
  year={2007},
  publisher={Springer}
}

@article{madronich1983measurement,
  title={Measurement of the photodissociation coefficient of {NO2} in the atmosphere: I. Method and surface measurements},
  author={Madronich, S and Hastie, DR and Ridley, BA and Schiff, HI},
  journal={Journal of Atmospheric Chemistry},
  volume={1},
  number={1},
  pages={3--25},
  year={1983},
  publisher={Springer}
}

@article{shpitser2008complete,
  author  = {Ilya Shpitser and Judea Pearl},
  title   = {Complete Identification Methods for the Causal Hierarchy},
  journal = {Journal of Machine Learning Research},
  year    = {2008},
  volume  = {9},
  pages   = {1941--1979}
}

@article{xie2026synthesising,
  title={Synthesising textual-and model-based evidence for proactive urban policymaking: A case in air quality management},
  author={Xie, Xiang and Kassem, Mohamad and Cabaneros, Sheen Mclean and Pan, Haixiao},
  journal={Journal of Environmental Management},
  volume={404},
  pages={129470},
  year={2026},
  publisher={Elsevier}
}

@article{kelp2022online,
  title={An online-learned neural network chemical solver for stable long-term global simulations of atmospheric chemistry},
  author={Kelp, Makoto M and Jacob, Daniel J and Lin, Haipeng and Sulprizio, Melissa P},
  journal={Journal of Advances in Modeling Earth Systems},
  volume={14},
  number={6},
  pages={e2021MS002926},
  year={2022},
  publisher={Wiley Online Library}
}

@article{bodnar2025foundation,
  title={A foundation model for the Earth system},
  author={Bodnar, Cristian and Bruinsma, Wessel P and Lucic, Ana and Stanley, Megan and Allen, Anna and Brandstetter, Johannes and Garvan, Patrick and Riechert, Maik and Weyn, Jonathan A and Dong, Haiyu and others},
  journal={Nature},
  volume={641},
  number={8065},
  pages={1180--1187},
  year={2025},
  publisher={Nature Publishing Group UK London}
}

@article{park2023linear,
  title={The linear representation hypothesis and the geometry of large language models},
  author={Park, Kiho and Choe, Yo Joong and Veitch, Victor},
  journal={arXiv preprint arXiv:2311.03658},
  year={2023}
}

@article{templeton2026scaling,
  title={Scaling monosemanticity: Extracting interpretable features from claude 3 sonnet},
  author={Templeton, Adly and Conerly, Tom and Marcus, Jonathan and Lindsey, Jack and Bricken, Trenton and Chen, Brian and Pearce, Adam and Citro, Craig and Ameisen, Emmanuel and Jones, Andy and others},
  journal={arXiv preprint arXiv:2605.29358},
  year={2026}
}

@article{shi2021abrupt,
  title={Abrupt but smaller than expected changes in surface air quality attributable to {COVID-19} lockdowns},
  author={Shi, Zongbo and Song, Congbo and Liu, Bowen and Lu, Gongda and Xu, Jingsha and Van Vu, Tuan and Elliott, Robert JR and Li, Weijun and Bloss, William J and Harrison, Roy M},
  journal={Science advances},
  volume={7},
  number={3},
  pages={eabd6696},
  year={2021},
  publisher={American Association for the Advancement of Science}
}

@article{hoel2013quantifying,
  title={Quantifying causal emergence shows that macro can beat micro},
  author={Hoel, Erik P and Albantakis, Larissa and Tononi, Giulio},
  journal={Proceedings of the National Academy of Sciences},
  volume={110},
  number={49},
  pages={19790--19795},
  year={2013},
  publisher={National Academy of Sciences}
}
